\documentclass[aps, prb,twocolumn]{revtex4-2}
\usepackage[english]{babel}
\usepackage{amsmath}
\usepackage{amssymb}
\usepackage{dsfont}
\usepackage{graphicx}
\usepackage{caption}
\usepackage{subcaption}
\usepackage[colorlinks=true, allcolors=blue]{hyperref}
\usepackage{manfnt}
\usepackage{tikz}
\usepackage{tikz-cd} 
\newcommand{\Z}{\mathbb{Z}}

\newcommand{\C}{\mathbb{C}}
\newcommand{\id}{\mathds{1}}
\newcommand{\ket}[1]{\vert #1 \rangle}

\usetikzlibrary{3d}
\usetikzlibrary{decorations.markings,calc}
\usetikzlibrary{decorations.pathmorphing}
\usetikzlibrary{perspective}
\usetikzlibrary{snakes}
\usetikzlibrary{shapes.geometric}

\tikzset{
  pics/cube/.style={
  code={
\draw[ black,fill=red,opacity=0.7] {[canvas is xy plane at z=0.5] (0,0) rectangle (0.4,0.4)};
\draw[ black,fill=red,opacity=0.7] {[canvas is yz plane at x=0.4] (0,0) rectangle (0.4,0.5)};
\draw[ black,fill=red,opacity=0.7] {[canvas is xz plane at y=0.4] (0,0) rectangle (0.4,0.5)};
}}}

\begin{document}


\title{Dipoles and Anyonic Directional Confinement via Twisted Toric Codes}

\author{Jos\'e Garre Rubio}
\affiliation{ {\small Instituto de F\'isica Te\'orica, UAM/CSIC, C. Nicol\'as Cabrera 13-15, Cantoblanco, 28049 Madrid, Spain}}

\begin{abstract}
We introduce a modified 2D toric code Hamiltonian that exhibits explicit anyon confinement along a single spatial direction. By bounding the motion of these confined anyons, we obtain dipolar excitations with restricted mobility. We analyze the resulting logical operators, whose existence depends on the system size, as well as the structure of gapped boundaries and a tensor network representation of the ground state. Furthermore, when confinement is enforced in both directions, fractal-like excitations emerge, resulting in unpaired logical operators. 

We extend our construction to 3D models, such as the surface code and the $X$-cube model, leading to novel dipole-loop and dipole-planon excitations that arise from bounding confined excitations. These modifications are implemented through group cohomological twistings—projective representations of finite groups—with most examples based on $\mathbb{Z}_2 \times \mathbb{Z}_2$.
\end{abstract}

\maketitle

\section{Introduction}

Topological quantum error correction is one of the most promising approaches to fault-tolerant quantum computation, with the toric code serving as a paradigmatic example of how quantum information can be protected through topological order \cite{Kitaev03}. In these systems, quantum information is encoded in the ground state degeneracy of gapped Hamiltonians, where logical operations correspond to nontrivial loops commuting with all local terms. The robustness of this encoding stems from the energy gap, which protects against local perturbations, while the topological nature ensures that errors must generate excitations whose energy cost scales with system size.

Central to the physics of topological codes are anyonic excitations—quasiparticles created when the ground state is locally violated. In the standard toric code, anyons are deconfined: they can move freely without incurring additional energy cost. Their braiding statistics encode logical operations, making the controlled manipulation of anyons essential for topological quantum computation \cite{Preskill04}. However, this unrestricted mobility can also pose challenges for certain applications, motivating the exploration of mechanisms to control anyon dynamics while preserving topological protection.

Traditionally, anyon confinement—where the energy required to separate a pair of excitations grows with their distance—has been associated with phase transitions or the breakdown of topological order \cite{Bais02,Vidal_2009,IqbalSuch21}. More recently, fracton phases \cite{Vijay16} have demonstrated that controlled confinement and restricted mobility can enhance the capabilities of stabilizer codes \cite{Brown20,Song22}. This shift in perspective opens the door to engineering topological phases with tailored excitation dynamics and richer phenomenology.

In this work, we present a novel framework for introducing directional confinement of anyons in topological codes by breaking the vertex-plaquette and rotational symmetries of the toric code. We show that modifying the Hamiltonian allows for confinement along a chosen direction, which in turn alters the types of gapped boundaries \cite{Kitaev12} and changes the structure of logical operators.

The confinement can be lifted via two mechanisms. First, by binding left- and right-handed anyons—where this emergent chirality arises naturally from the modified Hamiltonian—we obtain dipolar excitations with constrained mobility. Second, by dressing the confined anyon with different anyon-type operators, we construct a deconfined dyon-like excitation. This latter process is only possible when the confined anyon operators have even support, leading to a system-size dependence in the existence of corresponding logical operators. This phenomenon relates to the notion of 'topological frustration' studied in Ref.~\cite{chen2025}.

\begin{figure}[h!]
     \centering
     \begin{subfigure}[b]{0.23\textwidth}
         \centering
         \includegraphics[width=\textwidth]{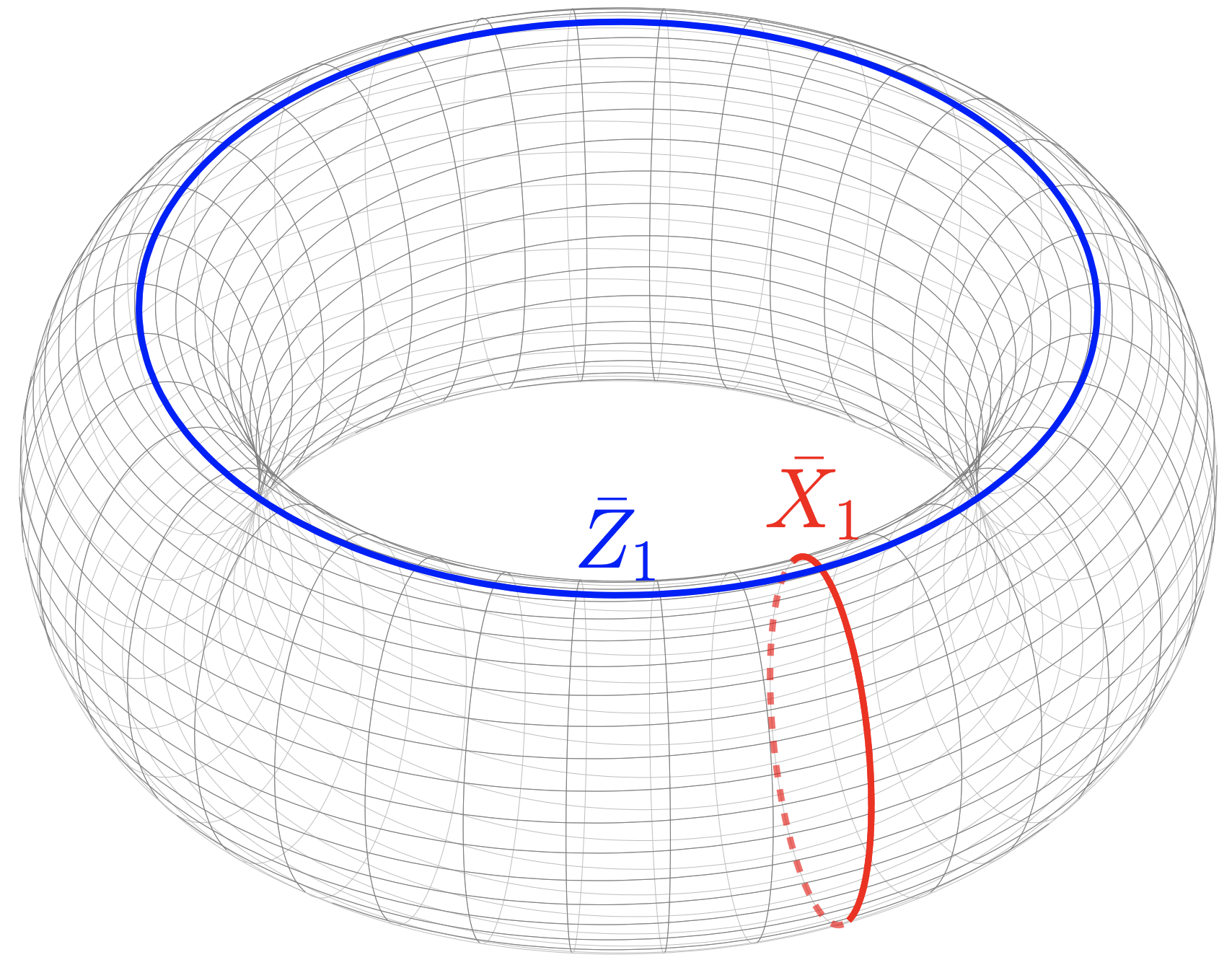}
     \end{subfigure}
     \hfill
     \begin{subfigure}[b]{0.23\textwidth}
         \centering
         \includegraphics[width=\textwidth]{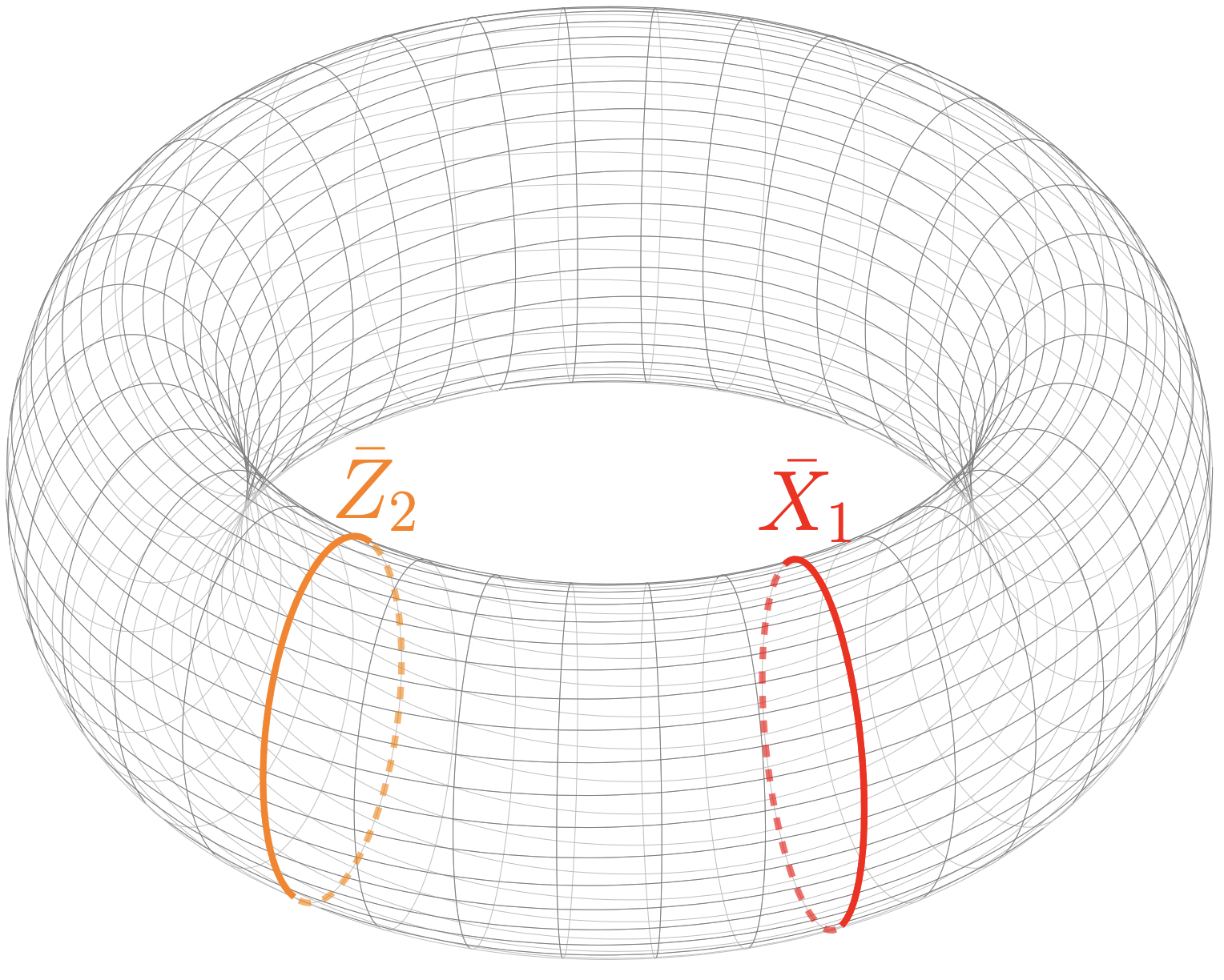}
     \end{subfigure}
\caption{Two types of ground state encodings found in our models. Left: a single pair of logical operators remains, encoding only half of the original ground space. Right: two uncoupled logical operators that do not encode qubits, yielding a classical-like code with topological protection.}
\label{fig:2graphs}
\end{figure}

Consequently, the effective topological order of the model depends on the system size, with certain logical operators being selectively eliminated. This transforms the original two-qubit code into a one-qubit code—or even into a classical-like code that nonetheless retains topological protection, as illustrated in Fig.~\ref{fig:2graphs}.

Extending this framework to three dimensions reveals even richer physics. Confined anyons form extended dipolar excitations, including loop-like and planon-like structures. These dipole-loops and dipole-planons exhibit hybrid mobility: they can move freely within specific subspaces while remaining constrained in orthogonal directions.

The structure of the paper is as follows: In Section~\ref{sec:background}, we provide the necessary mathematical background for constructing our models, focusing primarily on the regular projective representation of $\mathbb{Z}_2 \times \mathbb{Z}_2$ and its dual. Section~\ref{sec:mainmodel} introduces our main model, highlighting its novel features, including decorated Wilson loops, logical operators, anyonic excitations, gapped boundaries, and a tensor network representation of the ground state. In Section~\ref{sec:generalization}, we extend the model by introducing twists along the other spatial direction, modifying both plaquette and vertex terms, and generalizing the construction to arbitrary finite groups. Finally, Section~\ref{sec:3dmodels} discusses the application of our twisting framework to three-dimensional models, specifically the surface code and the $X$-cube model for the group $\mathbb{Z}_2 \times \mathbb{Z}_2$.

\section{The basic math ingredients}\label{sec:background}

We will work with the finite abelian group $\Z_2\times \Z_2 \equiv \Z^2_2$ generated by $a,b$ of order two. The regular representation, acting on $\C[\Z_2\times \Z_2]=\{\ket{g},g\in \Z^2_2 \}\simeq \C^{\otimes 4}$, is given by $X_g\ket{h}=\ket{gh}$, where in this case $X^\dagger_g = X_g $. The regular projective representation is given by
\begin{equation}
X^\alpha_a = \begin{pmatrix}
0&1&0&0\\
1&0&0&0\\
0&0&0&-i\\
0&0&i&0\\
\end{pmatrix} \ , 
X^\alpha_b = \begin{pmatrix}
0&0&1&0\\
0&0&0&i\\
1&0&0&0\\
0&-i&0&0\\
\end{pmatrix} \ ,
\end{equation}
and $X^\alpha_{ab}= -i\cdot X^\alpha_a\cdot  X^\alpha_b$. Since it is projective it satisfies that $X^\alpha_g X^\alpha_h = \alpha(g,h)X^\alpha_{gh}$ for all $ g,h\in \Z^2_2$. The phase factor $\alpha$ defined by: 

\begin{table}[h!]
\begin{tabular}{|c|c|c|c|c|}
\hline
$ \alpha(\downarrow,\rightarrow)$&  $e$& $a$ & $b$ & $ab$ \\ \hline
 $e$& $1$  & $1$ & $1$ & $1$ \\ \hline
 $a$& $1$ & $1$ & $i$ & $-i$ \\ \hline
 $b$&  $1$& $-i$ & $1$ & $i$ \\ \hline
 $ab$&  $1$& $i$ & $-i$ & $1$ \\ \hline
\end{tabular}
\end{table}
is a 2-cocycle \footnote{The set of 2-cocycles are classified by the second cohomology group, $\mathcal{H}^2[G,U(1)]$, which is abelian.} satisfying $\alpha(gh,k)\alpha(g,h)=\alpha(g,hk)\alpha(h,k)$. This implies that if $g$ and $h$ are non-trivial and different, their projective representations anticommute: 
\begin{equation}
    X^\alpha_g X^\alpha_h = - X^\alpha_h X^\alpha_g \ , \ e\neq g\neq h\neq e \ .
\end{equation} 
We will use also its conjugate projective representation 
$\overline{X^\alpha_g} \equiv X^{\bar{\alpha}}_g$ such that $X^{\bar{\alpha}}_g \otimes X^{{\alpha}}_g $ is linear and they commute:
\begin{equation}
    [X^\alpha_g ,X^{\bar{\alpha}}_h] = 0 \ , \ \forall g,h\in \Z^2_2 \ .
\end{equation}
We further need to define a representation for the dual group on $\C[\Z_2\times \Z_2]$ given by $Z_\chi\ket{g}=\chi(g)\ket{g}$, where $\chi$ is an irrep of $\Z^2_2$. Importantly, the operators defined on this space satisfy the followoing relation:
\begin{equation}\label{rel1}
 Z_\chi\cdot X^\gamma_g = \chi(g) \cdot X^\gamma_g \cdot Z_\chi \ ,
 \end{equation}
where $\gamma$ can be $\alpha$ or the trivial one. A useful way to labeling the non-trivial irreps is by assigning $\chi\equiv\hat{g}\neq 1$ if $\chi(g)=1$. Then, one can write \begin{equation}
    X^{\bar{\alpha}}_g X^{{\alpha}}_g = Z_{\hat{g}} \ .
\end{equation}
With such notation the irrep table reads
\begin{table}[h!]
\begin{tabular}{|c|c|c|c|c|}
\hline
$ \chi \backslash g $&  $e$& $a$ & $b$ & $ab$ \\ \hline
 $1$& $1$  & $1$ & $1$ & $1$ \\ \hline
 $\hat{a}$& $1$ & $1$ & $-1$ & $-1$ \\ \hline
 $\hat{b}$&  $1$& $-1$ & $1$ & $-1$ \\ \hline
 $\hat{ab}$&  $1$& $-1$ & $-1$ & $1$ \\ \hline
\end{tabular}
\end{table}
Finally, we can also construct a projective representation of the dual group. Since we are working with an abelian group, $G=\Z_2\times \Z_2$, its dual $\hat{G}$ is isomorphic to itself so there exists an isomorphism $\phi:\hat{G}\to G$.  Then, we can define a non-trivial 2-cocycle of $\hat{G}$ from $\alpha$ by $\beta \equiv \alpha \circ (\phi\otimes \phi)$. For convenience we will choose $\phi(\hat{g})=g$. 
Let us define the following projective representation of the dual group:
$$Z^\beta_{\hat{g}} = V \cdot X^\alpha_g \cdot V \ , \ V = \frac{1}{2} \begin{pmatrix}
1&1&1&1\\
1&1&-1&-1\\
1&-1&1&-1\\
1&-1&-1&1\\
\end{pmatrix}\ ,$$
where also $X_g = V\cdot Z_{\hat{g}}\cdot V $, so that $Z^\beta_{\chi} \cdot Z^\beta_{\chi'} = \beta(\chi,\chi')\cdot Z^\beta_{\chi\chi'}$ and 
\begin{equation}
 Z^\sigma_\chi\cdot X_g = \chi(g)\cdot X_g \cdot Z^\sigma_\chi \ ,
 \end{equation}
where $\sigma=\beta,1$. Moreover
\begin{equation}
    Z^{\bar{\beta}}_{\hat{g}} Z^{{\beta}}_{\hat{g}} = X_{g} \ .
\end{equation}
At last, all necessary components are in place to construct twisted Hamiltonians built upon $\mathbb{Z}_2 \times \mathbb{Z}_2$.

\section{Twisting the toric code by a 2-cocycle}\label{sec:mainmodel}

The Hamiltonian we want to study is a deformation of the regular $\Z_2\times \Z_2$-toric code where we {\it twist} the plaquette terms by a 2-cocycle:
\begin{equation}\label{Halpha}
 H^\alpha_{\Z_2^2 TC} = -\sum_{g,\chi}
\begin{tikzpicture}[baseline=0cm]
\draw[very thick, gray] (0.5,-0.5) rectangle (1.5,0.5);
\node[] at (0.5,0) {$X^{\bar{\alpha}}_{g}$};
\node[] at (1.5,0) {$X^\alpha_{{g}}$};
\node[] at (1,0.5) {$X_{{g}}$};
\node[] at (1,-0.5) {$X_{{g}}$};
\end{tikzpicture}
+
\begin{tikzpicture}[baseline=0cm]
\draw[very thick, gray] (0.2,0)--++(1.5,0);
\draw[very thick, gray] (1,-0.7)--++(0,1.5);
\node[] at (0.5,0) {$Z_{\chi}$};
\node[] at (1.5,0) {$Z_{\chi}$};
\node[] at (1,0.5) {$Z_{\chi}$};
\node[] at (1,-0.5) {$Z_{\chi}$};
\end{tikzpicture} \ .
\end{equation}

This Hamiltonian is commuting and frustration free since can be seen as a stabilizer code, where its ground state construction can be found in Ref.\cite{Garre24Emergent}. The vertex term is the Kitaev one\cite{Kitaev03}: it enforces the product of group elements in the vertex to be trivial since $\sum_\chi \chi(g)\bar{\chi}(h) = \delta_{h,g} $ for abelian groups. The introduction of the non-trivial $\alpha$, through the projective representations $X_g^\alpha$ and $X_g^{\bar{\alpha}}$, has severe consequences for the topological order of the Hamiltonian. 

We first notice that the Wilson loops, the 1-form symmetry, enforced by the plaquette Hamiltonian terms are decorated with $Z$-matrices:
$$
\begin{tikzpicture}[baseline=1.5cm,scale=0.75]
\draw[step=1.0,gray,very thick,xshift=0.5cm,yshift=0.5cm] (-0.5,-0.5) grid (3.5,3.5);
\draw[red, opacity=0.6, very thick,rounded corners] (0.5,0.5) rectangle (3.5,3.5);
\foreach \y in {1,2,3}{
\node[] at (\y,3.5) {$X_{g}$};
\node[] at (\y,0.5) {$X_{g}$};
\node[] at (0.5,\y) {$X^{\bar{\alpha}}_{g}$};
\node[] at (1.5,\y) {$Z_{\hat{g}}$};
\node[] at (2.5,\y) {$Z_{\hat{g}}$};
\node[] at (3.5,\y) {$X^{{\alpha}}_{g}$};
}
\draw[purple, dotted, opacity=0.6, thick] (0.5,2)--++ (3,0);
\end{tikzpicture}
=
\begin{tikzpicture}[baseline=1.5cm,scale=0.75]
\draw[step=1.0,gray,very thick,xshift=0.5cm,yshift=0.5cm] (-0.5,-0.5) grid (3.5,3.5);
\draw[red, opacity=0.6, very thick,rounded corners] (0.5,0.5) rectangle (3.5,3.5);
\foreach \y in {1,2,3}{
\node[] at (\y,3.5) {$X_{g}$};
\node[] at (\y,0.5) {$X_{g}$};
\node[] at (0.5,\y) {$X^{\bar{\alpha}}_{g}$};
\node[] at (3.5,\y) {$X^{{\alpha}}_{g}$};
}
\node[] at (1,2.5) {$Z_{\hat{g}}$};
\node[] at (3,2.5) {$Z_{\hat{g}}$};
\node[] at (1.5,1) {$Z_{\hat{g}}$};
\node[] at (2.5,1) {$Z_{\hat{g}}$};
\end{tikzpicture}
\ ,
$$
where we have chosen a loop with an odd number of vertical plaquettes and the equality is given by the application of vertex Hamiltonian terms. These Wilson loops for an even number of vertical plaquettes are simpler since only $Z$-decoration on the vertical direction is needed:
$$
\begin{tikzpicture}[baseline=1.5cm,scale=0.6]
\draw[step=1.0,gray,very thick,xshift=0.5cm,yshift=0.5cm] (-0.5,-0.5) grid (3.5,4.5);
\draw[red, opacity=0.6, very thick,rounded corners] (0.5,0.5) rectangle (3.5,4.5);
\foreach \y in {1,2,3,4}{
\node[] at (0.5,\y) {$X^{\bar{\alpha}}_{g}$};
\node[] at (3.5,\y) {$X^{{\alpha}}_{g}$};
}
\foreach \x in {1,2,3}{
\node[] at (\x,4.5) {$X_{g}$};
\node[] at (\x,0.5) {$X_{g}$};
}
\node[] at (1,3.5) {$Z_{\hat{g}}$};
\node[] at (3,3.5) {$Z_{\hat{g}}$};
\node[] at (1,1.5) {$Z_{\hat{g}}$};
\node[] at (3,1.5) {$Z_{\hat{g}}$};
\end{tikzpicture}
$$

Let us study the logical operators commuting with the Hamiltonian in the torus. The strings constructed by a horizontal loop of $X_g$-operators and a vertical loop of $Z_\chi$-operators, acting both on horizontal edges, are the logical operators that we denote by $\bar{X}_1$ and $\bar{Z}_1$ (let us omit the $g,\chi$ labels for the sake of simplicity). Naively this suggests that these operators act in the encoded subspace of the ground space of $H^\alpha_{\Z_2^2 TC}$ similarly to the untwisted case, although this needs a further treatment.

The logical operators $\bar{Z}_1$ can be deformed freely by acting with the vertex Hamiltonian terms. $\bar{X}_1$ is deformed by the plaquette and vertex terms, introducing $Z$-operators decorating the vertical strings, in the following way:
$$
\begin{tikzpicture}[baseline=0cm]
\draw[step=1.0,gray,very thick,xshift=0.5cm,yshift=0.5cm] (-0.5,0.5) grid (6.5,3.5);
\draw[red, opacity=0.6, very thick,rounded corners] (0,1.5)--++ (2.5,0)--++(0,2)--++(3,0)--++(0,-2)--++(1.5,0);
\node[] at (1,1.5) {$X_{g}$};
\node[] at (2,1.5) {$X_{g}$};
\node[] at (2.5,2) {$X^{\bar{\alpha}}_{g}$};
\node[] at (2.5,3) {$X^{\bar{\alpha}}_{g}$};
\node[] at (5.5,2) {$X^{{\alpha}}_{g}$};
\node[] at (5.5,3) {$X^{{\alpha}}_{g}$};
\node[] at (3,2.5) {$Z_{\hat{g}}$};
\node[] at (5,2.5) {$Z_{\hat{g}}$};
\node[] at (3,3.5) {$X_{g}$};
\node[] at (4,3.5) {$X_{g}$};
\node[] at (5,3.5) {$X_{g}$};
\node[] at (6,1.5) {$X_{g}$};
\end{tikzpicture}
$$

We denote by $\bar{Z}_2$ the logical operator corresponding to a horizontal loop composed of $Z_\chi$-operators on the vertical edges which is deformable without any decoration.

The decoration of $\bar{X}_1$ showcases the fact that this logical operator and $\bar{Z}_2$ are linked by the action of the plaquette Hamiltonian terms: $\bar{X}_1$ moves to the adjacent row plus a $\bar{Z}_2$ action. This also implies, using the action of the vertex terms, that $\bar{X}_1$ moves between non-consecutive rows:
$$
\begin{tikzpicture}[baseline=0cm]
\draw[step=1.0,gray,very thick,xshift=0.5cm,yshift=0.5cm] (-0.2,-0.5) grid (4.2,2.5);
\draw[red, opacity=0.6, very thick] (0.3,2.5)--++ (4.4,0);
\draw[red, opacity=0.6, very thick] (0.3,1.5)--++ (4.4,0);
\draw[red, opacity=0.6, very thick] (0.3,0.5)--++ (4.4,0);
\draw[red, opacity=0.6, very thick,dotted] (0.3,2)--++ (4.4,0);
\foreach \x in {1,2,3,4}{
\node[] at (\x,2.5) {$X_{g}$};
\node[] at (\x,1.5) {$X_{g}$};
\node[] at (\x,0.5) {$X_{g}$};
\node[] at (\x-0.5,2) {$Z_{\hat{g}}$};
}
\node[] at (4.5,2) {$Z_{\hat{g}}$};
 \draw[thick,|->] (5,2.5) to [out=0,in=0] (4.8,1.5);
 \draw[thick,->] (5,2.5) to [out=0,in=0] (4.8,2);
 \draw[thick,|->] (0.2,2.5) to [out=180,in=180] (0.2,0.5);
\begin{scope}[shift={(5.5,2)}]
\node[] at (0,0) {\large$\prod$};
\draw[very thick, gray] (0.5,-0.5) rectangle (1.5,0.5);
\node[] at (0.5,0) {$X^{\bar{\alpha}}_{g}$};
\node[] at (1.5,0) {$X^\alpha_{{g}}$};
\node[] at (1,0.5) {$X_{{g}}$};
\node[] at (1,-0.5) {$X_{g}$};
\end{scope}
\end{tikzpicture}
$$
For system sizes with an odd number of rows, the product of all plaquette terms for a given group element results in $\bar{Z}_2$. This means that $\bar{Z}_2$ does not correspond to a logical operator for those system sizes. 

For an even number of rows, $\bar{X}_1$ can move freely, using the action of the Hamiltonian terms, between even (or odd) rows and can jump from even to odd (and vice-versa) adding $\bar{Z}_2$ so $\bar{X}^{odd}_1 \cdot \bar{X}^{even}_1 =  \bar{Z}_2$ or equivalently:
$$
\prod_{row}
\begin{tikzpicture}[baseline=0cm]
\draw[very thick, gray] (0.5,-0.5) rectangle (1.5,0.5);
\node[] at (0.5,0) {$X^{\bar{\alpha}}_{g}$};
\node[] at (1.5,0) {$X^\alpha_{{g}}$};
\node[] at (1,0.5) {$X_{{g}}$};
\node[] at (1,-0.5) {$X_{{g}}$};
\end{tikzpicture}
= \bar{X}^{odd}_1(g) \cdot \bar{Z}_2(\hat{g}) \cdot \bar{X}^{even}_1(g) \ .
$$

Let us now consider the result of applying the plaquette terms in the whole torus, with total number of vertical plaquettes $p_y^{in}+p_y^{out}$, except in a contractible rectangular patch with $p_y^{in}$ vertical plaquettes. If $p_y^{out}$ is odd the product will result in $\bar{Z}_1$, similarly if $p_y^{in}$ is odd there will be a non-contractible horizontal $Z$-loop connecting the vertical boundaries of the rectangular patch. For $p_y^{in}$ and $p_y^{out}$ even, the result is simply:
$$
\begin{tikzpicture}[baseline=1.5cm,scale=0.6]
\draw[step=1.0,gray,very thick,xshift=0.5cm,yshift=0.5cm] (-0.75,-0.5) grid (3.75,4.5);
\draw[red, opacity=0.6, very thick,rounded corners] (0.5,0.5) rectangle (3.5,4.5);
\foreach \y in {1,2,3,4}{
\node[] at (3.5,\y) {$X^{\bar{\alpha}}_{g}$};
\node[] at (0.5,\y) {$X^{{\alpha}}_{g}$};
}
\foreach \x in {1,2,3}{
\node[] at (\x,4.5) {$X_{g}$};
\node[] at (\x,0.5) {$X_{g}$};
}
\node[] at (0,3.5) {$Z_{\hat{g}}$};
\node[] at (4,3.5) {$Z_{\hat{g}}$};
\node[] at (0,1.5) {$Z_{\hat{g}}$};
\node[] at (4,1.5) {$Z_{\hat{g}}$};
\end{tikzpicture}
$$
whose multiplication with the corresponding Wilson loop of the rectangular patch results in a product of vertex terms.

The existence of the logical operator partner of $\bar{Z}_2$ also depends on the numbers of plaquettes in the vertical direction. If this number is even then a vertical loops of $X_g^\alpha$ (or $X_g^{\bar{\alpha}}$) decorated with $Z_{\hat{g}}$ corresponds to a logical operator that we denote by $\bar{Y}_2$ (since they are associated with the dyons). Similar to the product of Wilson loops, the two aparent distinct flavors of $\bar{Y}_2$, given by $X^\alpha$ and $X^{\bar{\alpha}}$, are connected by a product of vertex terms:

$$
\begin{tikzpicture}[baseline=1.5cm,scale=0.6]
\draw[step=1.0,gray,very thick,xshift=0.5cm,yshift=0.5cm] (-0.5,-0.5) grid (0.75,4.5);
\draw[red, opacity=0.6, very thick] (0.5,0)--++ (0,5);
\foreach \y in {1,2,3,4}{
\node[] at (0.5,\y) {$X^{\bar{\alpha}}_{g}$};
}
\node[] at (1,1.5) {$Z_{\hat{g}}$};
\node[] at (1,3.5) {$Z_{\hat{g}}$};
\end{tikzpicture}
\times
\begin{tikzpicture}[baseline=1.5cm,scale=0.6]
\draw[step=1.0,gray,very thick,xshift=0.5cm,yshift=0.5cm] (-0.75,-0.5) grid (0.75,4.5);
\draw[red, opacity=0.6, very thick] (0.5,0)--++ (0,5);
\foreach \y in {1,2,3,4}{
\node[] at (0.5,\y) {$X^{{\alpha}}_{g}$};
}
\node[] at (0,3.5) {$Z_{\hat{g}}$};
\node[] at (0,1.5) {$Z_{\hat{g}}$};
\end{tikzpicture}
= \prod 
\begin{tikzpicture}[baseline=0cm,scale=0.6]
\draw[very thick, gray] (0.2,0)--++(1.5,0);
\draw[very thick, gray] (1,-0.7)--++(0,1.5);
\node[] at (0.5,0) {$Z_{\hat{g}}$};
\node[] at (1.5,0) {$Z_{\hat{g}}$};
\node[] at (1,0.5) {$Z_{\hat{g}}$};
\node[] at (1,-0.5) {$Z_{\hat{g}}$};
\end{tikzpicture}
$$
The $Z$-decoration on these logical operators can be placed on even or odd horizontal edges: $\bar{Y}^{odd}_2$ and $\bar{Y}^{even}_2$ which interact non-trivially with $\bar{X}^{odd}_1$ and $\bar{X}^{even}_1$ respectively. Moreover, $\bar{Y}^{odd}_2$ and $\bar{Y}^{even}_2$ multiply to $\bar{Z}_1$:
$$
\begin{tikzpicture}[baseline=1.5cm,scale=0.6]
\draw[step=1.0,gray,very thick,xshift=0.5cm,yshift=0.5cm] (-0.75,-0.5) grid (0.75,4.5);
\draw[red, opacity=0.6, very thick] (0.5,0)--++ (0,5);
\foreach \y in {1,2,3,4}{
\node[] at (0.5,\y) {$X^{{\alpha}}_{g}$};
}
\node[] at (0,3.5) {$Z_{\hat{g}}$};
\node[] at (0,1.5) {$Z_{\hat{g}}$};
\end{tikzpicture}
\times
\begin{tikzpicture}[baseline=1.5cm,scale=0.6]
\draw[step=1.0,gray,very thick,xshift=0.5cm,yshift=0.5cm] (-0.75,-0.5) grid (0.75,4.5);
\draw[red, opacity=0.6, very thick] (0,0)--++ (0,5);
\foreach \y in {0,1,2,3,4}{
\node[] at (0,\y+0.5) {$Z_{\hat{g}}$};
}
\end{tikzpicture}
= 
\begin{tikzpicture}[baseline=1.5cm,scale=0.6]
\draw[step=1.0,gray,very thick,xshift=0.5cm,yshift=0.5cm] (-0.75,-0.5) grid (0.75,4.5);
\draw[red, opacity=0.6, very thick] (0.5,0)--++ (0,5);
\foreach \y in {1,2,3,4}{
\node[] at (0.5,\y) {$X^{{\alpha}}_{g}$};
}
\node[] at (0,4.5) {$Z_{\hat{g}}$};
\node[] at (0,2.5) {$Z_{\hat{g}}$};
\node[] at (0,0.5) {$Z_{\hat{g}}$};
\end{tikzpicture}
$$

For system sizes with an odd number of plaquettes in the vertical direction there is no $\bar{Y}_2$ since there is always an unpair excitation (and neither $\bar{Z}_2$). As such, the only logical operators are $\bar{Z}_1,\bar{X}_1$, which correspond to a 4-fold ground state degeneracy in the torus, encoding just two qubits as the regular toric code. This phenomenon is akin to the 'topological frustration' studied in Ref.~\cite{chen2025}.

Therefore, for an even number of vertical plaquettes we have two pairs of logical operators $\{ \bar{Z}_1,\bar{X}^{odd}_1,\bar{X}^{even}_1\}$ and $\{ \bar{Z}_2, \bar{Y}^{odd}_2,\bar{Y}^{even}_2\}$ that interact non-trivially:
$$\bar{X}^{odd}_1 \cdot \bar{X}^{even}_1 =  \bar{Z}_2 \ , \quad \bar{Y}^{odd}_2 \cdot \bar{Y}^{even}_2 = \bar{Z}_1 $$

We emphasize that there is no logical operator associated with  vertical loops of $X$'s (neither $X_g,X^\alpha_g$ nor $X^{\bar{\alpha}}_g$) alone, without $Z$-decorations, acting on the vertical edges, since they do not commute with the Hamiltonian. Also we note that a doubled vertical loop of $X^\alpha_g\otimes X^{\bar{\alpha}}_g$ is the product of plaquette Hamiltonian terms so it does not correspond to a logical operator neither.

The same behavior is found if we would have twisted the vertical edges of the plaquette instead of the horizontal ones. In section \ref{sec:moretwist} we study the full twisting of the plaquette Hamiltonian terms and also the combination of twisted plaquettes and twisted vertex terms.

\subsection{Excitations}

The natural excitations of the toric code Hamiltonian are plaquette or vertex violations created by $Z$ and $X$ operators respectively. These string operators on top of the ground state create pairs of quasiparticles that are free to move; the so-called anyons. The introduction of the twisting $\alpha$ modifies that behavior by confining the plaquette excitations created by $X$-operators, i.e. the energy of a string of such operators is proportional to the string length.

The local operators ${Z_\chi,X_g,X^\alpha_g,X^{\bar{\alpha}}_{g}}$ excite in the following way the plaquettes and vertex terms, marked in red, of the Hamiltonian:
$$
\begin{tikzpicture}[baseline=0cm]
\draw[step=1.0,gray,very thick,xshift=0.5cm,yshift=0.5cm] (-0.5,-0.5) grid (7.5,4.5);
\node[] at (1.5,1) {$X^{\bar{\alpha}}_{g}$};
\draw[fill=red] (1.5,1.5) circle (0.1);
\draw[fill=red] (1.5,0.5) circle (0.1);
\draw[fill=red] (2,1) circle (0.1);

\node[] at (1.5,3) {$X^{{\alpha}}_{g}$};
\draw[fill=red] (1.5,3.5) circle (0.1);
\draw[fill=red] (1.5,2.5) circle (0.1);
\draw[fill=red] (1,3) circle (0.1);

\node[] at (3.5,4) {$Z_{\chi}$};
\draw[fill=red] (3,4) circle (0.1);
\draw[fill=red] (4,4) circle (0.1);

\node[] at (4,2.5) {$Z_{\chi}$};
\draw[fill=red] (4,2) circle (0.1);
\draw[fill=red] (4,3) circle (0.1);

\node[] at (4,0.5) {$X_{g}$};
\draw[fill=red] (4.5,0.5) circle (0.1);
\draw[fill=red] (3.5,0.5) circle (0.1);

\node[] at (5.5,4) {$X^{\bar{\alpha}}_{g}$};
\node[] at (6,3.5) {$Z_{\hat{g}}$};
\node[] at (5.5,3) {$X^{\bar{\alpha}}_{g}$};
\draw[fill=red] (5.5,4.5) circle (0.1);
\draw[fill=red] (5.5,2.5) circle (0.1);

\draw[fill=cyan, draw=blue, opacity=0.7] (7,1.5) ellipse (0.7 and 0.25);
\draw[fill=cyan, draw=blue, opacity=0.7] (7,0.5) ellipse (0.7 and 0.25);
\node[] at (7.5,1) {$X^{{\alpha}}_{g}$};
\node[] at (6.5,1) {$X^{\bar{\alpha}}_{g}$};
\draw[fill=red] (6.5,1.5) circle (0.1);
\draw[fill=red] (7.5,1.5) circle (0.1);
\draw[fill=red] (6.5,0.5) circle (0.1);
\draw[fill=red] (7.5,0.5) circle (0.1);

\end{tikzpicture} 
$$

\begin{figure}[h]
\includegraphics[width=0.5\textwidth]{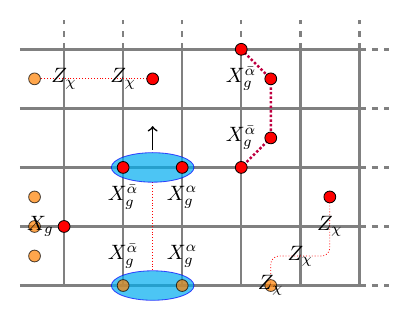}
\caption{The different excitations with their Hamiltonian term violation are shown. The red dots show the bulk violations and the orange dots shows the possible boundary excitations.}
\label{fig:exclat}
\end{figure}

The combination of the previous operators give rise to the following type of excitations, depicted in Fig.~\ref{fig:exclat}:
\begin{itemize}
    \item  A string of $Z_\chi$'s, in the dual lattice, creates plaquettes excitations at its endpoints. These are deconfined anyons, the charges
    .
    \item A horizontal string on the direct lattice of $X_g$-operators creates a pair of vertex excitations at its ends. This corresponds to fluxes. However, in order to bend this string and move it horizontally using $X^{{\alpha}}_{g}$ or $X^{\bar{\alpha}}_{g}$ a decoration with $Z_{\hat{g}}$ is needed. This decoration is possible for even string sizes otherwise an endpoint plaquette excitation is left. This decoration adds a dyonic character to this vertical string since it can interact non-trivially with a horizontal flux, see Sec.~\ref{sec:braid}.
    
    \item A vertical string of $X^{{\alpha}}_{g}$, on the direct lattice, creates vertex excitations at its ends and excites all plaquettes on the left of the string. The energy, then is proportional to the length of the string. Similarly, a $X^{{\alpha}}_{g}$-string is also confined exciting the right adjacent plaquettes. We say that the fluxes are confined vertically. However, we can bound these left and right oriented excitations in the next type.
    
    \item A vertical string of $X^{\bar{\alpha}}_{g}\otimes X^{{\alpha}}_{g}$, by bounding the previously discussed confined fluxes, creates a pair of dipoles at its ends. A dipole is restricted to move freely in the vertical direction. A dipole can be split in the horizontal direction by the action of $Z_{\hat{g}}$ but, this disables it to move vertically.
\end{itemize}

We note that a diagonal string composed of $X$-operators, any combination of ${X_g,X^\alpha_g,X^{\bar{\alpha}}_{g}}$, is not deconfined as seemly can be though by how these individual operators violate geometrically the Hamiltonian terms.

\subsection{Braiding}\label{sec:braid}

An important property of topologically ordered models are the braiding statistics of its anyonic excitations. Braiding is properly defined for deconfined anyons which are free to move, meaning that they can move unitarily and the energy of a pair is independent of its distance.

The braiding between a charge, constructed with $Z_\chi$ and a flux-dyon, constructed with horizontal $X_g$ and vertical $X_g^\alpha$ decorated with $Z_{\hat{g}}$, results in a phase factor of $\chi(g)$. This is akin to the braiding of charge and dyon (independent of its charge part) of the toric code.

However, the braiding between the composite flux-dyon excitation with itself is not topological in the sense that it depends on the actual geometric path. This is because the vertical decoration with $Z$-matrices acts only in half of the edges, as such, a horizontal string of $X_g$ can overlap or not with such decoration. For example, a configuration where the strings do not overlap is:
$$ 
\begin{tikzpicture}[baseline=3cm,scale=0.75]
\draw[step=1.0,gray,very thick,xshift=0.5cm,yshift=0.5cm] (-0.75,-0.5) grid (4.75,6.5);
\draw[red, opacity=0.6, very thick,rounded corners] (0.5,0.5) rectangle (4.5,4.5);
\foreach \y in {1,2,3,4}{
\node[] at (4.5,\y) {$X^{\bar{\alpha}}_{h}$};
\node[] at (0.5,\y) {$X^{{\alpha}}_{h}$};
}
\foreach \x in {1,2,3,4}{
\node[] at (\x,4.5) {$X_{h}$};
\node[] at (\x,0.5) {$X_{h}$};
}
\node[] at (0,3.5) {$Z_{\hat{h}}$};
\node[] at (5,3.5) {$Z_{\hat{h}}$};
\node[] at (0,1.5) {$Z_{\hat{h}}$};
\node[] at (5,1.5) {$Z_{\hat{h}}$};
\foreach \y in {3,4,5,6}{
\node[] at (2.5,\y) {$X^{\bar{\alpha}}_{g}$};}
\node[] at (2,3.5) {$Z_{\hat{g}}$};
\node[] at (2,5.5) {$Z_{\hat{g}}$};
\draw[fill=red] (2.5,2.5) circle (0.1);
\draw[fill=red] (2.5,6.5) circle (0.1);
\end{tikzpicture}
=
\begin{tikzpicture}[baseline=3cm,scale=0.75]
\draw[step=1.0,gray,very thick,xshift=0.5cm,yshift=0.5cm] (1.25,1.5) grid (2.75,6.5);
\foreach \y in {3,4,5,6}{
\node[] at (2.5,\y) {$X^{\bar{\alpha}}_{g}$};}
\node[] at (2,3.5) {$Z_{\hat{g}}$};
\node[] at (2,5.5) {$Z_{\hat{g}}$};
\draw[fill=red] (2.5,2.5) circle (0.1);
\draw[fill=red] (2.5,6.5) circle (0.1);
\end{tikzpicture}
$$

In the next braiding configuration the overlap gives a non-trivial phase factor:
$$ 
\begin{tikzpicture}[baseline=3cm,scale=0.75]
\draw[step=1.0,gray,very thick,xshift=0.5cm,yshift=0.5cm] (-0.75,0.5) grid (4.75,6.5);
\draw[red, opacity=0.6, very thick,rounded corners] (0.5,1.5) rectangle (4.5,5.5);
\foreach \y in {2,3,4,5}{
\node[] at (4.5,\y) {$X^{\bar{\alpha}}_{h}$};
\node[] at (0.5,\y) {$X^{{\alpha}}_{h}$};
}
\foreach \x in {1,3,4}{
\node[] at (\x,5.5) {$X_{h}$};
\node[] at (\x,1.5) {$X_{h}$};
}
\node[] at (2,1.5) {$X_{h}$};
\node[] at (0,4.5) {$Z_{\hat{h}}$};
\node[] at (5,4.5) {$Z_{\hat{h}}$};
\node[] at (0,2.5) {$Z_{\hat{h}}$};
\node[] at (5,2.5) {$Z_{\hat{h}}$};
\foreach \y in {3,4,5,6}{
\node[] at (2.5,\y) {$X^{\bar{\alpha}}_{g}$};}
\node[] at (2,3.5) {$Z_{\hat{g}}$};
\node[] at (2,5.5) {$X_hZ_{\hat{g}}$};
\draw[fill=red] (2.5,2.5) circle (0.1);
\draw[fill=red] (2.5,6.5) circle (0.1);
\end{tikzpicture}
=
\hat{g}(h)
\begin{tikzpicture}[baseline=3cm,scale=0.75]
\draw[step=1.0,gray,very thick,xshift=0.5cm,yshift=0.5cm] (1.25,1.5) grid (2.75,6.5);
\foreach \y in {3,4,5,6}{
\node[] at (2.5,\y) {$X^{\bar{\alpha}}_{g}$};}
\node[] at (2,3.5) {$Z_{\hat{g}}$};
\node[] at (2,5.5) {$Z_{\hat{g}}$};
\draw[fill=red] (2.5,2.5) circle (0.1);
\draw[fill=red] (2.5,6.5) circle (0.1);
\end{tikzpicture}
$$

Here we see again a sublattice splitting, as in the case of the logical operators, that determines the behavior of the model. 

We finally note that dipoles have trivial braiding with $Z_\chi$-strings since the string of those is 'doubled' so they commute: $[Z_\chi\otimes Z_\chi, X^{\bar{\alpha}}_{g}\otimes X^{{\alpha}}_{g}]=0$.

\subsection{Boundary conditions}

In this section we extend the previous Hamiltonian to surface codes by proposing boundary Hamiltonian terms for the different gapped boundaries and we identify what anyon can condense there. We find that there is a difference between vertical and horizontal boundaries and also among the smooth and rough boundary types. The former is due to the directional confinement and the later due to the asymmetry of the twisting between plaquette and vertex terms.  

The different gapped boundaries of quantum doubles of $G$ are classified by the pairs $({H}\subseteq G,\gamma\in \mathcal{H}^2[H,U(1)])$ \cite{Kitaev12}, where $\mathcal{H}^2[H,U(1)]$ is the second cohomology group of $H$ classifying 2-cocycles. Due to the twist in the Hamiltonian, this classification changes, in particular, some of those classes fuses to the same one.

We first study the rough horizontal (left) boundary. In this case the six types of gapped boundaries depend on the pair $({H}\subseteq \Z_2^2,\gamma\in \mathcal{H}^2[H,U(1)])$: 
\begin{equation}
    H^{rough}_{h\!{\tiny-}\!\rm bdry}({H},\gamma)= -\sum_{g \in H}
\begin{tikzpicture}[baseline=-0.5cm]
\draw[very thick, gray] (-1,0)--(0,0)--(0,-1)--(-1,-1);
\draw[very thick, gray,dashed] (0,0.5)--(0,0)--(0.5,0);
\draw[very thick, gray,dashed] (0,-1.5)--(0,-1)--(0.5,-1);
\node[] at (-0.5,0) {$X^\gamma_{{g}}$};
\node[] at (0,-0.5) {$X^\alpha_{{g}}$};
\node[] at (-0.5,-1) {$X^{\bar{\gamma}}_{{g}}$};
\end{tikzpicture}
-
\sum_{\chi \in \widehat{G/H}}
\begin{tikzpicture}[baseline=0cm]
\draw[very thick, gray] (0,0)--++(-1,0);
\draw[very thick, gray] (0,-0.7)--++(0,1.4);
\draw[very thick, gray,dashed] (0,0)--(0.4,0);
\node[] at (-0.5,0) {$Z_{\chi}$};
\end{tikzpicture} 
\end{equation}
and the choice of $({H},\gamma)$ determines which anyons condense at the boundary:
\begin{itemize}
    \item $(\Z_2\times \Z_2, \alpha\neq 1)$ Both type of anyons, $X_g$ and $Z_\chi$, violate boundary Hamiltonian terms. As such, no anyon is condensed at the boundary.
    \item $(\Z_2\times \Z_2, 1)$ All $X_g$ can condense at the boundary: a string terminating in the edge commutes with all Hamiltonian terms.
    \item $(\Z_2, 1)$ The three different cases are given by $g\neq e$ with $H=\langle g \rangle$ so $Z_{\hat{g}}$ and $X_g$ are condensed at the boundary.
    \item $(\Z_1, 1)$ All $Z_\chi$ can condensed at the edge.
\end{itemize}

For the smooth horizontal (left) boundary, the possible boundary Hamiltonian only depends on $\hat{H}\subseteq \widehat{\Z_2^2}$. This is because a twist on the truncated vertex terms, introducing a $Z^\beta_\chi$-operator, would not commute with the plaquette bulk terms. Then, the form of these boundary Hamiltonian are
\begin{equation}
    H^{smooth}_{h\!{\tiny-}\!\rm bdry}(\hat{H})= 
-\sum_{\chi \in \hat{H}}
\begin{tikzpicture}[baseline=-0.3cm]
\draw[very thick, gray] (0,0)--++(1,0);
\draw[very thick, gray] (0,-1) rectangle (1,1);
\draw[very thick, gray,dashed] (1,1.4)--(1,1)--(1.4,1);
\draw[very thick, gray,dashed] (1,0)--(1.4,0);
\draw[very thick, gray,dashed] (1,-1.4)--(1,-1)--(1.4,-1);
\draw[very thick, gray,dashed] (0,-1)--(0,-1.4);
\draw[very thick, gray,dashed] (0,1)--(0,1.4);
\node[] at (0,-0.5) {$Z_{\chi}$};
\node[] at (0,0.5) {$Z_{\chi}$};
\node[] at (0.5,0) {$Z_{\chi}$};
\end{tikzpicture} 
-\sum_{g | \hat{H}(g)=1}
\begin{tikzpicture}[baseline=0.3cm]
\draw[very thick, gray] (0,0) rectangle (1,1);
\draw[very thick, gray,dashed] (0,-0.4)--(0,1.4);
\draw[very thick, gray,dashed] (1,-0.4)--(1,1.4);
\draw[very thick, gray,dashed] (1,1)--(1.4,1);
\draw[very thick, gray,dashed] (1,0)--(1.4,0);
\node[] at (0,0.5) {$X^{\alpha}_{{g}}$};
\end{tikzpicture}
\end{equation}
So the different anyon condensation patterns are:
\begin{itemize}
    \item $(\Z_2\times \Z_2)$ All $Z_\chi$ can condensed at the boundary.
    \item $(\Z_2)$ The three different cases are given by $\chi \neq 1$ with $\hat{H}=\langle \hat{g} \rangle$ so $Z_{\hat{g}}$ and $X_g$ are condensed at the boundary.
    \item $(\Z_1)$ All $X_g$ can condensed at the boundary.
\end{itemize}

We now turn our attention to vertical boundaries. The excitations that can end here are $Z_\chi$ strings, vertically decorated $X_g^\alpha$ strings and dipoles. 

Let us focus first on the smooth vertical (bottom) boundaries. We can construct a boundary Hamiltonian for all pairs $(\hat{H}\subseteq \widehat{\Z_2^2},\beta\in \mathcal{H}^2[\hat{H},U(1)])$:

\begin{align}
H^{smooth}_{v\!{\tiny-}\!\rm bdry}(\hat{H},\beta)= 
& -\sum_{\chi \in \hat{H}}
\begin{tikzpicture}[baseline=0.3cm]
\draw[very thick, gray] (1,0)--++(0,1);
\draw[very thick, gray] (0,0) rectangle (2,1);
\draw[very thick, gray,dashed] (-0.4,0)--(2.4,0);
\draw[very thick, gray,dashed] (-0.4,1)--(2.4,1);
\draw[very thick, gray,dashed] (0,1)--(0,1.4);
\draw[very thick, gray,dashed] (1,1)--(1,1.4);
\draw[very thick, gray,dashed] (2,1)--(2,1.4);
\node[] at (1.5,0) {$Z^\beta_{\chi}$};
\node[] at (1,0.5) {$Z_{\chi}$};
\node[] at (0.5,0) {$Z^{\bar{\beta}}_{\chi}$};
\end{tikzpicture} 
\\ &
-\sum_{g | \hat{H}(g)=1}
\begin{tikzpicture}[baseline=0.3cm]
\draw[very thick, gray] (0,0) rectangle (1,1);
\draw[very thick, gray,dashed] (-0.4,1)--(1.4,1);
\draw[very thick, gray,dashed] (-0.4,0)--(1.4,0);
\draw[very thick, gray,dashed] (0,1)--(0,1.4);
\draw[very thick, gray,dashed] (1,1)--(1,1.4);
\node[] at (0.5,0) {$X_{{g}}$};
\end{tikzpicture} \notag
\end{align}

\begin{itemize}
    \item For any of the pairs $(\hat{H},\beta)$ a string of $Z_\chi$ ending on a vertical edge commute with these boundary Hamiltonians, not creating an excitation. However, if the $Z_\chi$ string is terminated in the ending horizontal edge, it is condensed only if $\chi \in \hat{H}$ and $\beta=1$.
    \item For every $\hat{H}$ the vertical decorated $X_g^\alpha$ string, condenses at the boundary if $\hat{H}(g)=1$.
    \item Dipoles are never condensed at the boundary, they always excite some boundary Hamiltonian terms.
\end{itemize}

Finally, the rough vertical (bottom) boundary depend on the pair $({H}\subseteq \Z_2^2,\gamma\in \mathcal{H}^2[H,U(1)])$:

\begin{equation}
H^{rough}_{v\!{\tiny-}\!\rm bdry}(H,\gamma)= -\sum_{g \in H}
\begin{tikzpicture}[baseline=-0.3cm]
\draw[very thick, gray] (0,-1) --++ (0,1)--++(1,0)--++(0,-1);
\node[] at (1,-0.5) {$X^\gamma_{g}$};
\node[] at (0,-0.5) {$X^{\bar{\gamma}}_{g}$};
\draw[very thick, gray,dashed] (-0.4,0)--(1.4,0);
\draw[very thick, gray,dashed] (0,0)--(0,0.4);
\draw[very thick, gray,dashed] (1,0)--(1,0.4);
\node[] at (0.5,0) {$X_{g}$};
\end{tikzpicture} 
-
\sum_{\chi \in \widehat{G/H}}
\begin{tikzpicture}[baseline=-0.3cm]
\draw[very thick, gray] (0.3,0) --++ (1.4,0);
\draw[very thick, gray] (1,0) --++ (0,-1);
\draw[very thick, gray,dashed] (1,0)--(1,0.4);
\node[] at (1,-0.5) {$Z_{{\chi}}$};
\end{tikzpicture}    
\end{equation}
whose different anyon condensation pattern are:
\begin{itemize}
    \item $(\Z_2\times \Z_2, \alpha\neq 1)$ Only the dipoles can condense at the boundary.
    \item $(\Z_2\times \Z_2, 1)$ No anyon nor dipole can condense at the boundary.
    \item $(\Z_2, 1)$ The three different cases are given by $g\neq 1$ with $H=\langle g \rangle$ so $Z_{\hat{g}}$ can condensed at the boundary.
    \item $(\Z_1, 1)$ All $Z_\chi$ can condensed at the vertical boundary.
\end{itemize}
As such, the decorated flux cannot condense at this vertical boundary.

\begin{figure}[h]
  \includegraphics[width=0.5\textwidth]{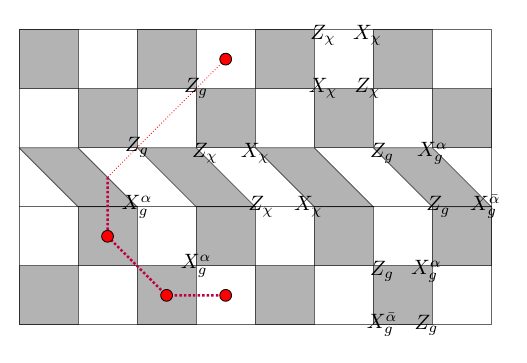}
\caption{Lattice dislocation defects permutes between anyons of different colored plaquettes as in Ref.~\cite{Bombin_2010}. However, since in our model they have different nature, the defects permute between deconfined and confined anyons. This representation of the model on the vertices of chessboard pattern is described in Ref.~\cite{Garre24Emergent}. All plaquette terms are shown in the picture, where we need to define the representations $X_\chi\ket{\sigma}=\ket{\chi \sigma}$ and $Z_g\ket{\chi}=\chi(g)\ket{\chi}$ on $\C[\hat{G}]$.}
\label{fig:dislocation}
\end{figure}

\vspace{1cm}
For the sake of completeness we show in Fig.~\ref{fig:dislocation} that if lattice defects are introduced, these permute between confined and unconfined anyons.

\subsection{PEPS representation}

Projected entangled pair states (PEPS) are a class of tensor network states very suitable for describing 2d gapped phases \cite{Verstraete04} in particular, topological order. 

The local property on the tensors that has been identified for describing ground states of topologically ordered systems is a purely virtual symmetry dubbed 'pulling through' \cite{Sahinoglu14}. The relevant class of PEPS that describe Kitaev quantum double models based on a group $G$ \cite{Kitaev03} are the so-called $G$-injective PEPS \cite{Schuch10}. These are defined as:
$$ 
\sum_g
\begin{tikzpicture}[baseline=0cm]
\draw[very thick, gray] (-0.25,0)--++(-0.5,0);
\draw[very thick, gray] (0.25,0)--++(0.5,0);
\draw[very thick, gray] (0,0.25)--++(0,0.5);
\draw[very thick, gray] (0,-0.25)--++(0,-0.5);
\draw[very thick] (0.25,0)--++(0.1,0.1);
\draw[very thick] (-0.25,0)--++(0.1,0.1);
\draw[very thick] (0,0.25)--++(0.1,0.1);
\draw[very thick] (0,-0.25)--++(0.1,0.1);

\node[] at (0.5,0) {$g$};
\node[] at (-0.5,0) {$\bar{g}$};
\node[] at (0,0.5) {$g$};
\node[] at (0,-0.5) {$\bar{g}$};
\end{tikzpicture} 
=
\begin{tikzpicture}[baseline=0cm]
\draw[very thick, gray] (-0.5,0)--++(1,0);
\draw[very thick, gray] (0,-0.5)--++(0,1);
\draw[very thick] (0,0)--++(0.2,0.2);
\filldraw[] (0,0) circle (0.1);
\end{tikzpicture} 
=
\begin{tikzpicture}[baseline=0cm]
\draw[very thick, gray] (-0.5,0)--++(1,0);
\draw[very thick, gray] (0,-0.5)--++(0,1);
\draw[very thick] (0,0)--++(0.2,0.2);
\filldraw[] (0,0) circle (0.1);
\node[] at (0.75,0) {$g$};
\node[] at (-0.75,0) {$\bar{g}$};
\node[] at (0,0.75) {$g$};
\node[] at (0,-0.75) {$\bar{g}$};
\end{tikzpicture} \ ,
$$
where $g$ and $\bar{g}$ denotes for simplicity the matrices representing $g$ and its inverse respectively.

To describe the ground state and the excitations of the  Hamiltonian model of Eq.~\eqref{Halpha}, we propose the following class of PEPS:
$$
\begin{tikzpicture}[baseline=0cm]
\draw[very thick, gray] (-0.5,0)--++(1,0);
\draw[very thick, gray] (0,-0.5)--++(0,1);
\draw[very thick] (0,0)--++(0.2,0.2);
\filldraw[] (0,0) circle (0.1);
\end{tikzpicture}
=
\sum_g
\begin{tikzpicture}[baseline=0cm]
\draw[very thick, gray] (-0.25,0)--++(-1,0);
\draw[very thick, gray] (0.25,0)--++(1,0);
\draw[very thick, gray] (0,0.25)--++(0,1);
\draw[very thick, gray] (0,-0.25)--++(0,-1);
\draw[very thick] (0.25,0)--++(0.1,0.1);
\draw[very thick] (-0.25,0)--++(0.1,0.1);
\draw[very thick] (0,0.25)--++(0.1,0.1);
\draw[very thick] (0,-0.25)--++(0.1,0.1);
\node[] at (0.75,0) {$X^\alpha_g$};
\node[] at (-0.75,0) {$X^{\bar{\alpha}}_g$};
\node[] at (0,0.75) {$X_g$};
\node[] at (0,-0.75) {$X_{g}$};
\end{tikzpicture} \ . 
$$
This tensor has the following virtual symmetries
$$
\begin{tikzpicture}[baseline=0cm]
\draw[very thick, gray] (-0.5,0)--++(1,0);
\draw[very thick, gray] (0,-0.5)--++(0,1);
\draw[very thick] (0,0)--++(0.2,0.2);
\node[] at (0.75,0) {$X^\alpha_g$};
\node[] at (-0.75,0) {$X^{\bar{\alpha}}_g$};
\node[] at (0,0.75) {$X_g$};
\node[] at (0,-0.75) {$X_{g}$};
\filldraw[] (0,0) circle (0.1);
\end{tikzpicture} 
=
\begin{tikzpicture}[baseline=0cm]
\draw[very thick, gray] (-0.5,0)--++(1,0);
\draw[very thick, gray] (0,-0.5)--++(0,1);
\draw[very thick] (0,0)--++(0.2,0.2);
\filldraw[] (0,0) circle (0.1);
\end{tikzpicture} 
\ , 
\begin{tikzpicture}[baseline=0cm]
\draw[very thick, gray] (-0.5,0)--++(1,0);
\draw[very thick, gray] (0,-0.5)--++(0,1);
\draw[very thick] (0,0)--++(0.2,0.2);
\node[] at (0,0.75) {$X_{g}$};
\filldraw[] (0,0) circle (0.1);
\end{tikzpicture} 
=
\begin{tikzpicture}[baseline=0cm]
\draw[very thick, gray] (-0.5,0)--++(1,0);
\draw[very thick, gray] (0,-0.5)--++(0,1);
\draw[very thick] (0,0)--++(0.2,0.2);
\node[] at (0.75,0) {$X^\alpha_g$};
\node[] at (-0.75,0) {$X^{\bar{\alpha}}_g$};
\node[] at (0,-0.75) {$X_g$};
\filldraw[] (0,0) circle (0.1);
\end{tikzpicture} \ ,
$$
where we can identify the last equation with the regular pulling through equation in the vertical direction. However, the pulling through relation in the horizontal direction works in a different way:
\begin{equation}\label{modPT}
\begin{tikzpicture}[baseline=0cm]
\draw[very thick, gray] (-0.5,0)--++(1,0);
\draw[very thick, gray] (0,-0.5)--++(0,1);
\draw[very thick] (0,0)--++(0.2,0.2);
\filldraw[] (0,0) circle (0.1);
\node[] at (-0.75,0) {$X^{\bar{\alpha}}_g$};
\end{tikzpicture} 
=
\begin{tikzpicture}[baseline=0cm]
\draw[very thick, gray] (-0.5,0)--++(1,0);
\draw[very thick, gray] (0,-0.5)--++(0,1);
\draw[very thick] (0,0)--++(0.2,0.2);
\draw[fill=yellow] (0,0) circle (0.1);
\node[] at (-0.2,-0.2) {$\hat{g}$};
\node[] at (0.75,0) {$X^\alpha_g$};
\node[] at (0,0.75) {$X_g$};
\node[] at (0,-0.75) {$X_{g}$};
\end{tikzpicture}
\end{equation}
where the new introduced yellow tensor is defined as follows:
$$
\begin{tikzpicture}[baseline=0cm]
\draw[very thick, gray] (-0.5,0)--++(1,0);
\draw[very thick, gray] (0,-0.5)--++(0,1);
\draw[very thick] (0,0)--++(0.2,0.2);
\draw[fill=yellow] (0,0) circle (0.1);
\node[] at (-0.2,-0.2) {$\chi$};
\end{tikzpicture}
=
\sum_g \chi(g)
\begin{tikzpicture}[baseline=0cm]
\draw[very thick, gray] (-0.25,0)--++(-0.75,0);
\draw[very thick, gray] (0.25,0)--++(0.75,0);
\draw[very thick, gray] (0,0.25)--++(0,0.75);
\draw[very thick, gray] (0,-0.25)--++(0,-0.75);
\draw[very thick] (0.25,0)--++(0.1,0.1);
\draw[very thick] (-0.25,0)--++(0.1,0.1);
\draw[very thick] (0,0.25)--++(0.1,0.1);
\draw[very thick] (0,-0.25)--++(0.1,0.1);
\node[] at (0.65,0) {$X^\alpha_g$};
\node[] at (-0.65,0) {$X^{\bar{\alpha}}_g$};
\node[] at (0,0.65) {$X_g$};
\node[] at (0,-0.65) {$X_{g}$};
\end{tikzpicture} 
$$
and it satisfies the following:
$$
\begin{tikzpicture}[baseline=0cm]
\draw[very thick, gray] (-0.5,0)--++(1,0);
\draw[very thick, gray] (0,-0.5)--++(0,1);
\draw[very thick] (0,0)--++(0.2,0.2);
\draw[fill=yellow] (0,0) circle (0.1);
\node[] at (-0.2,-0.2) {$\chi$};
\node[] at (-0.75,0) {$X^{\bar{\alpha}}_g$};
\node[] at (0.75,0) {$X^\alpha_g$};
\node[] at (0,0.75) {$X_g$};
\node[] at (0,-0.75) {$X_{g}$};
\end{tikzpicture} 
=
\chi(g)
\begin{tikzpicture}[baseline=0cm]
\draw[very thick, gray] (-0.5,0)--++(1,0);
\draw[very thick, gray] (0,-0.5)--++(0,1);
\draw[very thick] (0,0)--++(0.2,0.2);
\draw[fill=yellow] (0,0) circle (0.1);
\node[] at (-0.2,-0.2) {$\chi$};
\end{tikzpicture} 
$$
so that the tensor is charged: it transforms as the irrep $\chi$ under the action of the virtual symmetry $G$. The previous statement holds due to the following calculation:
$X^{\bar{\alpha}}_g\otimes \id (\sum_h X^{\bar{\alpha}}_h \otimes X^{{\alpha}}_h ) = \sum_h \bar{\alpha}(g,h)X^{\bar{\alpha}}_{gh} \otimes X^{{\alpha}}_h  = \sum_h \bar{\alpha}(g,hg)X^{\bar{\alpha}}_{h} \otimes \bar{\alpha}(h,g)X^{{\alpha}}_h X^{{\alpha}}_g =  (\sum_h \frac{\bar{\alpha}(h,g)}{\alpha(g,hg)}X^{\bar{\alpha}}_h \otimes X^{{\alpha}}_h ) \id \otimes X^{{\alpha}}_g $ where
it turns out that $\frac{\bar{\alpha}(h,g)}{\alpha(g,hg)}= \chi_{\hat{g}}(h)$.

We now argue that the modified pulling through of Eq.~\eqref{modPT} encapsulates correctly the confining properties of the anyons. 

In this representation deconfined fluxes are represented as a virtual horizontal string of $X_g$ operators, whose endpoint determine the plaquette excitations. The middle string is deformable, using the vertical pulling through in the following way:

$$
\begin{tikzpicture}[baseline=0cm]
\draw[step=1.0,gray,very thick,xshift=0.5cm,yshift=0.5cm,opacity=0.7] (-0.5,-0.5) grid (4.5,2.5);
\filldraw[red,thick] (1,1) circle (0.1)--(4,1) circle (0.1);
\draw[red, opacity=0.7,thick,dashed,rounded corners] (1,1)--++(0,0.5)--++(0.5,0.5)--++(0.5,-0.5)--++(0.5,0.5)--++(0.5,-0.5)--++(0.5,0.5)--++(0.5,-0.5)--++(0,-0.5);

\foreach \y in {1,2,3}{
\node[] at (\y+0.5,1) {$X_{g}$};
\node[ opacity=0.7] at (\y+0.5,2) {$X_{g}$};

\foreach \x in {1,2,3,4,5}{
\draw[very thick] (\x-0.5,\y-0.5)--++(0.2,0.2);
\filldraw[] (\x-0.5,\y-0.5) circle (0.1);
}}
\node[opacity=0.7] at (1,1.5) {$X^{\bar{\alpha}}_{g}$};
\node[opacity=0.7] at (2,1.5) {$Z_{\hat{g}}$};
\node[opacity=0.7] at (3,1.5) {$Z_{\hat{g}}$};
\node[opacity=0.7] at (4,1.5) {$X^{{\alpha}}_{g}$};

\end{tikzpicture} \ ,
$$
where the initial string is drawn in solid red and the equivalent deformed one is in dashed lines. Notice that the decoration with $Z_{\hat{g}}$-operators, not present in the regular untwisted case of $G$-injective PEPS though, does not correspond to plaquette excitations.

By using the twisted pulling through relation we can see that a $X^{\bar{\alpha}}_g$ excite its top and bottom plaquettes, like a flux, but also corresponds to a charge excitation (since the yellow tensor transforms as an irrep):
$$
\begin{tikzpicture}[baseline=0cm]

\filldraw[red, opacity=0.7] (-0.5,0.5) circle (0.1)--(-0.5,-0.5) circle (0.1);
\draw[very thick, gray] (-1,-1) rectangle (1,1);
\draw[very thick, gray] (-1.2,0)--++(2.4,0);
\draw[very thick, gray] (0,-1.2)--++(0,2.4);
\draw[very thick] (0,0)--++(0.2,0.2);
\filldraw[] (0,0) circle (0.1);
\draw[very thick] (-1,0)--++(0.2,0.2);
\filldraw[] (-1,0) circle (0.1);
\draw[very thick] (0,-1)--++(0.2,0.2);
\filldraw[] (0,-1) circle (0.1);
\draw[very thick] (0,1)--++(0.2,0.2);
\filldraw[] (0,1) circle (0.1);
\draw[very thick] (1,0)--++(0.2,0.2);
\filldraw[] (1,0) circle (0.1);
\node[] at (-0.5,0) {$X^{\bar{\alpha}}_g$};
\end{tikzpicture} 
=
\begin{tikzpicture}[baseline=0cm]
\draw[red, opacity=0.7,rounded corners] (-0.5,0.5) --++(1,0)--++ (0,-1)--++(-1,0) circle (0.1);
\filldraw[red, opacity=0.7] (-0.5,0.5) circle (0.1);
\filldraw[red, opacity=0.7] (-0.5,-0.5) circle (0.1);
\draw[very thick, gray] (-1,-1) rectangle (1,1);
\draw[very thick, gray] (-1.2,0)--++(2.4,0);
\draw[very thick, gray] (0,-1.2)--++(0,2.4);
\draw[very thick] (0,0)--++(0.2,0.2);
\draw[fill=yellow] (0,0) circle (0.1);
\draw[very thick] (-1,0)--++(0.2,0.2);
\filldraw[] (-1,0) circle (0.1);
\draw[very thick] (0,-1)--++(0.2,0.2);
\filldraw[] (0,-1) circle (0.1);
\draw[very thick] (0,1)--++(0.2,0.2);
\filldraw[] (0,1) circle (0.1);
\draw[very thick] (1,0)--++(0.2,0.2);
\filldraw[] (1,0) circle (0.1);
\node[] at (0,0.5) {$X_{g}$};
\node[] at (0,-0.5) {$X_g$};
\node[] at (0.5,0) {$X^\alpha_g$};
\end{tikzpicture} 
$$
where the red dots denotes the plaquette excitations and the string connects them through the operators that can be deformed via the twisted pulling through equation. Notice that the string cannot be moved further by pushing $X^\alpha_g$ to the right, that is a singularity of the directional confinement.

To see how the creation of dipoles through the bound state $X^\alpha_g \otimes X^{\bar{\alpha}}_g$ is reflected in the virtual level we just need to use the pulling through equation that results in:
$$
\begin{tikzpicture}[baseline=0cm]

\filldraw[red, opacity=0.7] (-0.5,0.5) circle (0.1)--(-0.5,-0.5) circle (0.1);
\filldraw[red, opacity=0.7] (0.5,0.5) circle (0.1)--(0.5,-0.5) circle (0.1);
\draw[very thick, gray] (-1,-1) rectangle (1,1);
\draw[very thick, gray] (-1.2,0)--++(2.4,0);
\draw[very thick, gray] (0,-1.2)--++(0,2.4);
\draw[very thick] (0,0)--++(0.2,0.2);
\filldraw[] (0,0) circle (0.1);
\draw[very thick] (-1,0)--++(0.2,0.2);
\filldraw[] (-1,0) circle (0.1);
\draw[very thick] (0,-1)--++(0.2,0.2);
\filldraw[] (0,-1) circle (0.1);
\draw[very thick] (0,1)--++(0.2,0.2);
\filldraw[] (0,1) circle (0.1);
\draw[very thick] (1,0)--++(0.2,0.2);
\filldraw[] (1,0) circle (0.1);
\node[] at (0.5,0) {$X^\alpha_g$};
\node[] at (-0.5,0) {$X^{\bar{\alpha}}_g$};
\end{tikzpicture} 
=
\begin{tikzpicture}[baseline=0cm]
\filldraw[red, opacity=0.7] (-0.5,0.5) circle (0.1)--(0.5,0.5) circle (0.1);
\filldraw[red, opacity=0.7] (-0.5,-0.5) circle (0.1)--(0.5,-0.5) circle (0.1);
\draw[very thick, gray] (-1,-1) rectangle (1,1);
\draw[very thick, gray] (-1.2,0)--++(2.4,0);
\draw[very thick, gray] (0,-1.2)--++(0,2.4);
\draw[very thick] (0,0)--++(0.2,0.2);
\filldraw[] (0,0) circle (0.1);
\draw[very thick] (-1,0)--++(0.2,0.2);
\filldraw[] (-1,0) circle (0.1);
\draw[very thick] (0,-1)--++(0.2,0.2);
\filldraw[] (0,-1) circle (0.1);
\draw[very thick] (0,1)--++(0.2,0.2);
\filldraw[] (0,1) circle (0.1);
\draw[very thick] (1,0)--++(0.2,0.2);
\filldraw[] (1,0) circle (0.1);
\node[] at (0,0.5) {$X_{g}$};
\node[] at (0,-0.5) {$X_g$};
\end{tikzpicture} 
$$
where the red dots and lines denotes from which operators the excitations are coming from. Then, the virtual operator $X^\alpha_g \otimes X^{\bar{\alpha}}_g$ placed horizontally is equivalent to $X_g \otimes X_g$ placed vertically whose string can be deformed freely and the excited plaquettes can be moved without creating more excitations. We left for future work the study of the ground state degeneracy using this PEPS representation.

\section{Generalizations}\label{sec:generalization}

\subsection{Twisting the other direction}
\label{sec:moretwist}

We can twist all the operators of the plaquette terms of the Hamiltonian in the following way:
$$ H^{\alpha,\alpha}_{\Z_2^2 TC} = -\sum
\begin{tikzpicture}[baseline=0cm]
\draw[very thick, gray] (0.5,-0.5) rectangle (1.5,0.5);
\node[] at (0.5,0) {$X^{\bar{\alpha}}_{g}$};
\node[] at (1.5,0) {$X^\alpha_{{g}}$};
\node[] at (1,0.5) {$X^\alpha_{{g}}$};
\node[] at (1,-0.5) {$X^{\bar{\alpha}}_{g}$};
\end{tikzpicture}
+
\begin{tikzpicture}[baseline=0cm]
\draw[very thick, gray] (0.2,0)--++(1.5,0);
\draw[very thick, gray] (1,-0.7)--++(0,1.5);
\node[] at (0.5,0) {$Z_{\chi}$};
\node[] at (1.5,0) {$Z_{\chi}$};
\node[] at (1,0.5) {$Z_{\chi}$};
\node[] at (1,-0.5) {$Z_{\chi}$};
\end{tikzpicture} 
$$
Then, the Wilson loop associated with the plaquettes are also decorated on the horizontal strings (after applying vertex operators):
$$ 
\begin{tikzpicture}[baseline=3cm,scale=0.75]
\draw[step=1.0,gray,very thick,xshift=0.5cm,yshift=0.5cm] (-0.5,0.5) grid (4.5,5.5);
\draw[red, opacity=0.6, very thick,rounded corners] (0.5,1.5) rectangle (4.5,5.5);
\foreach \y in {2,3,4,5}{
\node[] at (4.5,\y) {$X^{{\alpha}}_{g}$};
\node[] at (0.5,\y) {$X^{\bar{\alpha}}_{g}$};
\foreach \x in {1.5,2.5,3.5}{
\node[] at (\x,\y) {$Z_{\hat{g}}$};
\node[] at (\x,\y) {$Z_{\hat{g}}$};
}}
\foreach \x in {1,2,3,4}{
\node[] at (\x,5.5) {$X^{{\alpha}}_{g}$};
\node[] at (\x,1.5) {$X^{\bar{\alpha}}_{g}$};
\foreach \y in {2.5,3.5,4.5}{
\node[] at (\x,\y) {$Z_{\hat{g}}$};
\node[] at (\x,\y) {$Z_{\hat{g}}$};
}}
\end{tikzpicture}
=
\begin{tikzpicture}[baseline=3cm,scale=0.75]
\draw[step=1.0,gray,very thick,xshift=0.5cm,yshift=0.5cm] (-0.5,0.5) grid (4.5,5.5);
\draw[red, opacity=0.6, very thick,rounded corners] (0.5,1.5) rectangle (4.5,5.5);
\foreach \y in {2,3,4,5}{
\node[] at (4.5,\y) {$X^{{\alpha}}_{g}$};
\node[] at (0.5,\y) {$X^{\bar{\alpha}}_{g}$};
}
\node[] at (1,3.5) {$Z_{\hat{g}}$};
\node[] at (4,3.5) {$Z_{\hat{g}}$};
\foreach \x in {1,2,3,4}{
\node[] at (\x,5.5) {$X^{{\alpha}}_{g}$};
\node[] at (\x,1.5) {$X^{\bar{\alpha}}_{g}$};
}
\node[] at (2.5,2) {$Z_{\hat{g}}$};
\node[] at (2.5,5) {$Z_{\hat{g}}$};
\end{tikzpicture}
$$

Let us denote by $\bar{Z}_{\hat{x}/\hat{y}}$ the horizontal/vertical loop of $Z$-operators acting on the vertical/horizontal edges of a torus. Let us consider a torus with $p_x$ horizontal plaquettes and $p_y$ vertical plaquettes, then the following relation is satisfied:
$$
\prod_{all}
\begin{tikzpicture}[baseline=0cm]
\draw[very thick, gray] (0.5,-0.5) rectangle (1.5,0.5);
\node[] at (0.5,0) {$X^{\bar{\alpha}}_{g}$};
\node[] at (1.5,0) {$X^\alpha_{{g}}$};
\node[] at (1,0.5) {$X^\alpha_{{g}}$};
\node[] at (1,-0.5) {$X^{\bar{\alpha}}_{g}$};
\end{tikzpicture}
= 
\bar{Z}_{\hat{x}}^{|p_y|} \cdot \bar{Z}_{\hat{y}}^{|p_x|} \ ,
$$
where $|p_{x/y}|= \{0,1\}$ is the parity of $p_{x/y}$. As such, for odd $p_{x/y}$ the loop $\bar{Z}_{\hat{y}/\hat{x}}$ is not a logical operator. Similarly to the previous model, we can construct the logical operators $\bar{Y}_{\hat{x}}$ and $\bar{Y}_{\hat{y}}$ if $p_{x}$ and $p_{y}$ are even respectively. Moreover, in that case also a splitting between 2 sublattices, odd and even, of the logical operators occurs and they satisfy the following:
$$\bar{Y}^{odd}_{\hat{x}} \cdot \bar{Y}^{even}_{\hat{x}} =  \bar{Z}_{\hat{x}} \ , $$
and similarly for $\hat{y}$. Surprisingly, for the case when $p_{x}$ and $p_{y}$ are both odd there is no logical operators that encode a ground state degeneracy.

The single operators composing the Hamiltonian terms acting on top of a ground state violate the Hamiltonian in the following way:

$$
\begin{tikzpicture}[baseline=0cm]
\draw[step=1.0,gray,very thick,xshift=0.5cm,yshift=0.5cm] (-0.5,-0.5) grid (7.5,4.5);
\node[] at (1.5,1) {$X^{\bar{\alpha}}_{g}$};
\draw[fill=red] (1.5,1.5) circle (0.1);
\draw[fill=red] (1.5,0.5) circle (0.1);
\draw[fill=red] (2,1) circle (0.1);

\node[] at (1.5,3) {$X^{{\alpha}}_{g}$};
\draw[fill=red] (1.5,3.5) circle (0.1);
\draw[fill=red] (1.5,2.5) circle (0.1);
\draw[fill=red] (1,3) circle (0.1);

\node[] at (3.5,4) {$Z_{\chi}$};
\draw[fill=red] (3,4) circle (0.1);
\draw[fill=red] (4,4) circle (0.1);

\node[] at (4,2.5) {$Z_{\chi}$};
\draw[fill=red] (4,2) circle (0.1);
\draw[fill=red] (4,3) circle (0.1);

\node[] at (6,3.5) {$X^{\bar{\alpha}}_{g}$};
\draw[fill=red] (6,4) circle (0.1);
\draw[fill=red] (6.5,3.5) circle (0.1);
\draw[fill=red] (5.5,3.5) circle (0.1);

\node[] at (6,1.5) {$X^{{\alpha}}_{g}$};
\draw[fill=red] (6,1) circle (0.1);
\draw[fill=red] (6.5,1.5) circle (0.1);
\draw[fill=red] (5.5,1.5) circle (0.1);
\end{tikzpicture} 
$$
Charge excitations created by $Z$-operators are deconfined. The excitations created only by $X$-operators on top of a ground state are confined in both directions: the energy of a string, on the direct lattice, is proportional to its length. However, by decorating the strings of $X^\alpha$ by $Z$-operators, deconfined fluxes/dyons can be created. Moreover, combining $X^{{\alpha}}_{g}$ and $X^{\bar{\alpha}}_{g}$ dipoles that are free to move, either horizontally or vertically, can be created. 
The braiding between the fluxes/dyons depend on the sublattice where they live similarly to the previous model.

\subsection{Twisting both Hamiltonian terms}

We can also twist the vertex terms by a 2-cocycle of the dual group formed by the irreps. Since we are working with an abelian group, its dual is isomorphic to itself so there is an isomorphism $\phi:\hat{G}\to G$ with whom we define $\beta = \alpha \circ (\phi\otimes \phi)$. For simplicity we choose $\phi(\hat{g})=g$. Then, we define the following Hamiltonian:

$$ H^{\alpha,\beta}_{\Z_2^2 TC} = -\sum_{g,\chi}
\begin{tikzpicture}[baseline=0cm]
\draw[very thick, gray] (0.5,-0.5) rectangle (1.5,0.5);
\node[] at (0.5,0) {$X^{\bar{\alpha}}_{g}$};
\node[] at (1.5,0) {$X^\alpha_{{g}}$};
\node[] at (1,0.5) {$X_{{g}}$};
\node[] at (1,-0.5) {$X_{{g}}$};
\end{tikzpicture}
+
\begin{tikzpicture}[baseline=0cm]
\draw[very thick, gray] (0.2,0)--++(1.5,0);
\draw[very thick, gray] (1,-0.7)--++(0,1.5);
\node[] at (0.5,0) {$Z^{\bar{\beta}}_{\chi}$};
\node[] at (1.5,0) {$Z^\beta_{\chi}$};
\node[] at (1,0.5) {$Z_{\chi}$};
\node[] at (1,-0.5) {$Z_{\chi}$};
\end{tikzpicture} 
$$
The relevant operators excite the plaquette and vertex terms in the following way:

$$
\begin{tikzpicture}[baseline=0cm]
\draw[step=1.0,gray,very thick,xshift=0.5cm,yshift=0.5cm] (-0.5,-0.5) grid (7.5,4.5);
\node[] at (1.5,1) {$X^{\bar{\alpha}}_{g}$};
\draw[fill=red] (1.5,1.5) circle (0.1);
\draw[fill=red] (1.5,0.5) circle (0.1);
\draw[fill=red] (2,1) circle (0.1);

\node[] at (1.5,3) {$X^{{\alpha}}_{g}$};
\draw[fill=red] (1.5,3.5) circle (0.1);
\draw[fill=red] (1.5,2.5) circle (0.1);
\draw[fill=red] (1,3) circle (0.1);

\node[] at (3.5,4) {$Z_{\chi}$};
\draw[fill=red] (3,4) circle (0.1);
\draw[fill=red] (4,4) circle (0.1);

\node[] at (4,2.5) {$Z^\beta_{\chi}$};
\draw[fill=red] (4,2) circle (0.1);
\draw[fill=red] (3.5,2.5) circle (0.1);
\draw[fill=red] (4,3) circle (0.1);

\node[] at (4,0.5) {$X_{g}$};
\draw[fill=red] (4.5,0.5) circle (0.1);
\draw[fill=red] (3.5,0.5) circle (0.1);

\node[] at (6,3.5) {$Z^{\bar{\beta}}_{\chi}$};
\draw[fill=red] (6,4) circle (0.1);
\draw[fill=red] (6,3) circle (0.1);
\draw[fill=red] (6.5,3.5) circle (0.1);

\end{tikzpicture} 
$$

On top of the dipole vertex excitations formed by bounding $X^{{\alpha}}_{g}$ and $X^{\bar{\alpha}}_{g}$, as in the previous models, we also have dipole plaquette excitations formed by bounding $Z^{{\beta}}_{\chi}$ and $Z^{\bar{\beta}}_{\chi}$. We notice that in this model there are no deconfined dyons, neither decorating fluxes with $Z$-operators nor decorating charges with $X$-operators.

Interestingly, the combination of $X^{{\alpha}}_{g}$ and $Z^\beta_{\hat{g}}$ in a fractal-shape operator, the Sierpinski triangle, create either plaquette or vertex excitations on its corners:
$$
\begin{tikzpicture}[baseline=0cm]
\draw[step=1.0,gray,very thick,xshift=0.5cm,yshift=0.5cm] (-0.5,-0.5) grid (7.5,5.5);

\draw[fill=red] (1,3) circle (0.1);
\node[] at (1.5,3) {$X^{{\alpha}}_{g}$};
\node[] at (2,2.5) {$Z^\beta_{\hat{g}}$};
\node[] at (2,3.5) {$Z^\beta_{\hat{g}}$};
\node[] at (2.5,4) {$X^{{\alpha}}_{g}$};
\node[] at (2.5,2) {$X^{{\alpha}}_{g}$};
\node[] at (3,2.5) {$Z^\beta_{\hat{g}}$};
\node[] at (3,3.5) {$Z^\beta_{\hat{g}}$};
\node[] at (3,1.5) {$Z^\beta_{\hat{g}}$};
\node[] at (3,4.5) {$Z^\beta_{\hat{g}}$};
\draw[fill=red] (3,5) circle (0.1);
\draw[fill=red] (3,1) circle (0.1);
\begin{scope}[shift={(3.5,0.5)}]
\draw[fill=red] (1,3) circle (0.1);
\node[] at (1.5,3) {$Z^{{\beta}}_{\hat{g}}$};
\node[] at (2,2.5) {$X^\alpha_{{g}}$};
\node[] at (2,3.5) {$X^\alpha_{{g}}$};
\node[] at (2.5,4) {$Z^{{\beta}}_{\hat{g}}$};
\node[] at (2.5,2) {$Z^{{\beta}}_{\hat{g}}$};
\node[] at (3,2.5) {$X^\alpha_{{g}}$};
\node[] at (3,3.5) {$X^\alpha_{{g}}$};
\node[] at (3,1.5) {$X^\alpha_{{g}}$};
\node[] at (3,4.5) {$X^\alpha_{{g}}$};
\draw[fill=red] (3,5) circle (0.1);
\draw[fill=red] (3,1) circle (0.1);
\end{scope}
\end{tikzpicture} 
$$
where we have used that $[X^\alpha_g\otimes Z^\beta_{\hat{g}}, Z_{\chi} \otimes Z^\beta_{\chi}]=0$, since $\chi(g)= \frac{\beta(\chi,\hat{g})}{\beta(\hat{g},\chi)}$, to compute the violations  of the vertex terms and similarly with $\alpha$ for the plaquette terms. This excitations are akin to the symmetry defect excitations of the quantum Newman-Moore model \cite{Newman_1999}.

The same can be done with the pair 
$X^{\bar{\alpha}}_{g},Z^{\bar{\beta}}_{\hat{g}}$, that creates a similar fractal-operator but vertically reflected. Then, we can combine both left and right Sierpinski triangles to create a bow-tie fractal operator. In general, this will create excitations on its boundary, with a minimum of 4 violations in its corners. Moreover, for system sizes with a number of horizontal plaquettes $p_x \ge 2(n-1)$ and $p_y = 2^n$, provided that $n\ge 2$ and vertical periodic boundary conditions, those operators become symmetries of the Hamiltonian and then of the ground space.

Let us call these operators by $\mathcal{F}^p_g$, where $p$ denotes the middle plaquette. Similarly we can construct the analogous operator $\mathcal{F}^v_\chi$ using the pair $X^{\bar{\alpha}}_{g}, Z^{\bar{\beta}}_{\hat{g}}$ where the middle vertex is $v$. As an example, we show a $\mathcal{F}^v_{\hat{g}}$ in blue and a $\mathcal{F}^p_g$ in black in a vertical PBC with $p_y =4$ corresponding to Hamiltonian symmetries: 
$$
\begin{tikzpicture}[baseline=0cm]
\draw[step=1.0,gray,very thick,xshift=0.5cm,yshift=0.5cm] (-3,1) grid (4,5);
\draw[very thick, dashed] (-2.5,1.5)--++(7,0);
\draw[very thick, dashed] (-2.5,5.5)--++(7,0);
\node[] at (1.5,3) {$X^{{\alpha}}_{g}$};
\node[] at (2,2.5) {$Z^\beta_{\hat{g}}$};
\node[] at (2,3.5) {$Z^\beta_{\hat{g}}$};
\node[] at (2.5,4) {$X^{{\alpha}}_{g}$};
\node[] at (2.5,2) {$X^{{\alpha}}_{g}$};
\node[] at (3,2.5) {$Z^\beta_{\hat{g}}$};
\node[] at (3,3.5) {$Z^\beta_{\hat{g}}$};
\node[] at (3,1.5) {$Z^\beta_{\hat{g}}$};
\node[] at (3,4.5) {$Z^\beta_{\hat{g}}$};
\node[] at (0.5,3) {$X^{\bar{\alpha}}_{g}$};
\node[] at (0,2.5) {$Z^{\bar{\beta}}_{\hat{g}}$};
\node[] at (0,3.5) {$Z^{\bar{\beta}}_{\hat{g}}$};
\node[] at (-0.5,4) {$X^{\bar{\alpha}}_{g}$};
\node[] at (-0.5,2) {$X^{\bar{\alpha}}_{g}$};
\node[] at (-1,2.5) {$Z^{\bar{\beta}}_{\hat{g}}$};
\node[] at (-1,3.5) {$Z^{\bar{\beta}}_{\hat{g}}$};
\node[] at (-1,1.5) {$Z^{\bar{\beta}}_{\hat{g}}$};
\node[] at (-1,4.5) {$Z^{\bar{\beta}}_{\hat{g}}$};

\begin{scope}[shift={(0.6,0.6)},blue,opacity=0.8]
\node[] at (1.5,3) {$Z^\beta_{\hat{g}}$};
\node[] at (2,2.5) {$X^{{\alpha}}_{g}$};
\node[] at (2,3.5) {$X^{{\alpha}}_{g}$};
\node[] at (2.5,4) {$Z^\beta_{\hat{g}}$};
\node[] at (2.5,2) {$Z^\beta_{\hat{g}}$};
\node[] at (3,2.5) {$X^{{\alpha}}_{g}$};
\node[] at (3,3.5) {$X^{{\alpha}}_{g}$};
\node[] at (3,1.5) {$X^{{\alpha}}_{g}$};
\node[] at (3,4.5) {$X^{{\alpha}}_{g}$};
\node[] at (0.5,3) {$Z^{\bar{\beta}}_{\hat{g}}$};
\node[] at (0,2.5) {$X^{\bar{\alpha}}_{g}$};
\node[] at (0,3.5) {$X^{\bar{\alpha}}_{g}$};
\node[] at (-0.5,4) {$Z^{\bar{\beta}}_{\hat{g}}$};
\node[] at (-0.5,2) {$Z^{\bar{\beta}}_{\hat{g}}$};
\node[] at (-1,2.5) {$X^{\bar{\alpha}}_{g}$};
\node[] at (-1,3.5) {$X^{\bar{\alpha}}_{g}$};
\node[] at (-1,1.5) {$X^{\bar{\alpha}}_{g}$};
\node[] at (-1,4.5) {$X^{\bar{\alpha}}_{g}$};
\end{scope}

\begin{scope}[shift={(1.1,0.1)},orange,opacity=0]
\node[] at (1.5,3) {$X^{{\alpha}}_{g}$};
\node[] at (2,2.5) {$Z^\beta_{\hat{g}}$};
\node[] at (2,3.5) {$Z^\beta_{\hat{g}}$};
\node[] at (2.5,4) {$X^{{\alpha}}_{g}$};
\node[] at (2.5,2) {$X^{{\alpha}}_{g}$};
\node[] at (3,2.5) {$Z^\beta_{\hat{g}}$};
\node[] at (3,3.5) {$Z^\beta_{\hat{g}}$};
\node[] at (3,1.5) {$Z^\beta_{\hat{g}}$};
\node[] at (3,4.5) {$Z^\beta_{\hat{g}}$};
\node[] at (0.5,3) {$X^{\bar{\alpha}}_{g}$};
\node[] at (0,2.5) {$Z^{\bar{\beta}}_{\hat{g}}$};
\node[] at (0,3.5) {$Z^{\bar{\beta}}_{\hat{g}}$};
\node[] at (-0.5,4) {$X^{\bar{\alpha}}_{g}$};
\node[] at (-0.5,2) {$X^{\bar{\alpha}}_{g}$};
\node[] at (-1,2.5) {$Z^{\bar{\beta}}_{\hat{g}}$};
\node[] at (-1,3.5) {$Z^{\bar{\beta}}_{\hat{g}}$};
\node[] at (-1,1.5) {$Z^{\bar{\beta}}_{\hat{g}}$};
\node[] at (-1,4.5) {$Z^{\bar{\beta}}_{\hat{g}}$};
\end{scope}
\end{tikzpicture} 
$$
Importantly, all these operators commute between themselves:
$$ [\mathcal{F}^p_g,\mathcal{F}^{p'}_h]=[\mathcal{F}^v_\chi,\mathcal{F}^{v'}_{\sigma}]=[\mathcal{F}^p_g,\mathcal{F}^{v}_\chi]=0 \ ,$$
since either the same operator intersects between itself an even number of times and/or the overlap is between operators with opposite cocycle that commute. The existence of fractal symmetries does not contradict the no-go theorem prohibiting fractal logical operators in 2D quantum codes \cite{Zhu-nofraclog22}: Since the fractal operators commute between themselves, the encoded subspace is classical.

At the same time, they commute with both horizontal $Z_\chi$ and $X_g$ loops on vertical edges and horizontal edges respectively, dubbed $\bar{Z}_{\hat{x}}$  and $\bar{X}_{\hat{x}}$ respectively. Notice that these operators are only logical if there is a even number of vertical plaquettes since, otherwise the product of all plaquette terms and the product of all vertex terms result in $\bar{Z}_{\hat{x}}$ and $\bar{X}_{\hat{x}}$ respectively. For system sizes without the fractal symmetries, and even number of vertical plaquettes, these are the only logical operators. 

Since these operators do not have partners to generate the Pauli group, twice since we are working with $\mathbb{Z}_2^2$, they do not encode qubits. They loose some quantumness to encode, topologically, a classical ground state (4-bits each).

\subsection{Twisting any abelian surface code.}

Let us consider an abelian group $G$ and one of its 2-cocycles $\alpha$. Then we can construct the following commuting Hamiltonian:

$$ H^{\alpha}_{G} = -\sum_{g,\chi}
\begin{tikzpicture}[baseline=0cm]
\draw[very thick, gray] (0.5,-0.5) rectangle (1.5,0.5);
\node[] at (0.5,0) {$\tilde{X}^{\bar{\alpha}}_{g}$};
\node[] at (1.5,0) {$X^\alpha_{{g}}$};
\node[] at (1,0.5) {$X^\dagger_{{g}}$};
\node[] at (1,-0.5) {$X_{{g}}$};
\end{tikzpicture}
+
\begin{tikzpicture}[baseline=0cm]
\draw[very thick, gray] (0.2,0)--++(1.5,0);
\draw[very thick, gray] (1,-0.7)--++(0,1.5);
\node[] at (0.5,0) {${Z}^\dagger_{\chi}$};
\node[] at (1.5,0) {$Z_{\chi}$};
\node[] at (1,0.5) {$Z_{{\chi}}$};
\node[] at (1,-0.5) {$Z^\dagger_{{\chi}}$};
\end{tikzpicture} 
$$
where we now have introduced, since $G$ is not necessarily order two, the projective representation
$$\tilde{X}^{\bar{\alpha}}_{g} \ket{h} = \bar{\alpha}(h g^{-1},g)\ket{h g^{-1}}$$
that commutes with $X^\alpha$, given by $X^\alpha_g \ket{h}=\alpha(g,h)\ket{gh}$, satisfying
$ Z_\chi \tilde{X}^\alpha_g = \chi(g^{-1}) \cdot \tilde{X}^\alpha_g  Z_\chi $. Importantly 
\begin{equation}\label{Xslant}
 X^\alpha_g\cdot \tilde{X}^{\bar{\alpha}}_{g}= Z_{\iota_{g}^\alpha}\ \ {\rm and} \ \ X^\alpha_g X^{\alpha}_{h} = \iota_{g}^\alpha(h) X^{\alpha}_{h} X^\alpha_g
 \end{equation}
where we have used the slant product defined by:
$$\iota_{g}^\alpha(\cdot)=\frac{\alpha(g,\cdot)}{\alpha(\cdot,g)}\in {\rm Irrep}(G) \ . $$

In general, not all irreps of $G$ can be recovered from the set $ \{ \iota_g^\alpha , g\in G \}$ given $\alpha$. It turns out \cite{blanik25} that this is a subgroup ${\rm Im}(\iota^\alpha)\equiv \mathcal{I} \subseteq \hat{G}$ characterized by the irreps $\chi$ such that $\chi(K_\alpha)=1$ where $K_\alpha =\{ g\in G \ | \ \iota_{g}^\alpha = \hat{1} \}$.

The anyons of this models follow simply by cutting open the logical operators. Let us analyze the structure of the logical operators.
The strings constructed by a horizontal loop of $X_g$-operators and a vertical loop of $Z_\chi$-operators, acting both on horizontal edges, are the logical operators that we denote by $\bar{X}_1(G)$ and $\bar{Z}_1(\hat{G})$.
All the horizontal loops of $Z_\chi$-operators on the vertical edges commute with the Hamiltonian. However, for an odd number of vertical plaquettes the product of plaquette terms for a given element $g\in G$ is $Z_{\iota_{g}^\alpha}$ which is trivial if $g\in K_\alpha $. Then, for those system sizes the only logical operators are $\bar{Z}_2(\mathcal{I})$. 
For all $g\in K_\alpha$ the vertical loops of $X^\alpha_g$ and $X^{\bar{\alpha}}_g$ on the vertical edges commute with the Hamiltonian due to Eq.~\eqref{Xslant}: these are the logical operators $\bar{X}_2(K_\alpha)$. 
For the elements $g\notin K_\alpha$, we need to decorate the vertical loops of $X^\alpha_g$ with $Z_{\iota_{g}^\alpha}$, corresponding to the logical operators $\bar{Y}_2(G\backslash K_\alpha)$ for an even number of vertical plaquettes.

As summary we have that for an even number of vertical plaquettes the logical operators are
$$\{\bar{X}_1(G),\bar{Z}_1(\hat{G})\} \cup \{\bar{X}_2(K_\alpha),\bar{Y}_2(G\backslash K_\alpha),\bar{Z}_2(\hat{G}) \}\ , 
$$
and for an odd number:
$$\{\bar{X}_1(G),\bar{Z}_1(\hat{G})\} \cup \{\bar{X}_2(K_\alpha),\bar{Z}_2(\mathcal{I}) \} 
$$
encoding just a qu-$|G|$-bit and a qu-$|K_\alpha|$-bit. 

\section{3D twisted Hamiltonians}\label{sec:3dmodels}

In this section we study how 3D stabilizer codes can also be twisted by a 2-cocycle. We show how this twisting confines the string-like excitations of the 3D toric code and the planons of X-cube code, both models  based on the group $\Z_2\times \Z_2$.

\subsection{Twisted 3D $\Z_2\times\Z_2 $ surface code}
Starting from the Hamiltonian of the 3D surface code based on the group $\Z_2^2$ we twist the vertex terms on the $\hat{z}$-direction by the 2-cocycle $\alpha$. The resulting Hamiltonian has this form:
\begin{align}
H^{\alpha}_{\Z_2^2 SC} = &
-\sum_g
\begin{tikzpicture}[baseline=0cm]
\draw[gray,thick] (0,0,-1.2)--(0,0,1.2);
\draw[gray,thick] (-0.9,0,0)--(0.9,0,0);
\draw[gray,thick] (0,-0.9,0)--(0,0.9,0);
\node[] at (0,0,-0.6) {\footnotesize $X^\alpha_{g}$};
\node[] at (0,0,0.6) {\footnotesize $X^{\bar{\alpha}}_{g}$};
\node[] at (0,0.6,0) {\footnotesize $X_{g}$};
\node[] at (0,-0.6,0) {\footnotesize $X_{g}$};
\node[] at (0.6,0,0) {\footnotesize $X_{g}$};
\node[] at (-0.6,0,0) {\footnotesize $X_{g}$};
\end{tikzpicture}
\\ & - \sum_\chi
\begin{tikzpicture}[baseline=0.5cm]
\draw[ thick, gray] (0,0) rectangle (0.8,0.8);
\node[] at (0.8,0.4,0) {\footnotesize $Z_{\chi}$};
\node[] at (0,0.4,0) {\footnotesize $Z_{\chi}$};
\node[] at (0.4,0,0) {\footnotesize $Z_{\chi}$};
\node[] at (0.4,0.8,0) {\footnotesize $Z_{\chi}$};
\end{tikzpicture} 
+\!
\begin{tikzpicture}[baseline=0.3cm]
\draw[ thick, gray]{[canvas is yz plane at x=0] (0,0) rectangle (0.8,1.5)};
\node[] at (0,0,0.65) {\footnotesize$Z_{\chi}$};
\node[] at (0,0.8,0.85) {\footnotesize$Z_{\chi}$};
\node[] at (0,0.3,0) {\footnotesize$Z_{\chi}$};
\node[] at (0,0.5,1.5) {\footnotesize$Z_{\chi}$};
\end{tikzpicture} 
\!+
\begin{tikzpicture}[baseline=-0.3cm]
\draw[ thick, gray]{[canvas is xz plane at y=0] (0,0) rectangle (0.8,1.5)};
\node[] at (0,0,0.75) {\footnotesize$Z_{\chi}$};
\node[] at (0.8,0,0.75) {\footnotesize$Z_{\chi}$};
\node[] at (0.4,0,0) {\footnotesize$Z_{\chi}$};
\node[] at (0.4,0,1.5) {\footnotesize$Z_{\chi}$};
\end{tikzpicture} \ .
\notag
\end{align}
The $X_g$ operators acting on $\hat{x}$ and $\hat{y}$-oriented edges violates the four adjacent face terms. As such, a membrane of these operators creates string-like excitation at its boundary: its energy is proportional to its perimeter. This is the same as the regular 3D toric code.

However, the operators $X^\alpha_g$ and $X^{\bar{\alpha}}_g$ acting on $\hat{z}$-oriented edges violate the four adjacent face terms plus an additional vertex term. This implies that a membrane of these operators creates a string-like excitation decorated with a surface-like excitation: the energy is proportional to the area plus its perimeter. Thus, we can say that this excitation is confined.

We represent the excitations of these operators as red dots placed in either vertices or faces:
$$
\begin{tikzpicture}[baseline=0cm]
\draw[gray, thick] (0,0,0.75)--++(0,0,-1.5)--++(0,1,0)--++(0,0,1.5)--++(0,-1,0);
\draw[gray, thick] (0,0,0.75)--++(0,0,1.5)--++(0,1,0)--++(0,0,-1.5)--++(0,-1,0);
\draw[gray, thick] (0,0,0.75)--++(1,0,0)--++(0,1,0)--++(-1,0,0)--++(0,-1,0);
\draw[gray, thick] (0,0,0.75)--++(-1,0,0)--++(0,1,0)--++(1,0,0)--++(0,-1,0);
\draw[densely dotted, red,rounded corners] (0,0.5,0)--++(0.5,0.0)--++(0,0,1.5)--++(-1,0,0)--++(0,0,-1.5)--++(0.5,0,0);
\draw[fill=red]{[canvas is yz plane at x=0] (0.5,0) circle (0.075)};
\draw[fill=red]{[canvas is yz plane at x=0] (0.5,1.5) circle (0.075)};
\draw[fill=red]{[canvas is xy plane at z=0.75] (0.5,0.5) circle (0.075)};
\draw[fill=red]{[canvas is xy plane at z=0.75] (-0.5,0.5) circle (0.075)};
\node[] at (0,0.5,0.75) {$X_{g}$};
\end{tikzpicture} 
\ , \ 
\begin{tikzpicture}[baseline=0cm]
\draw[densely dotted, red,rounded corners] (-0.5,0,0)--++(0,0.5,0)--++(1,0,0)--++(0,-1,0)--++(-1,0,0)--(-0.5,0,0);
\draw[gray,thick] (0,0,-0.75)--++(0,0,1.5)--++(-1,0,0)--++(0,0,-1.5)--++(+1,0,0);
\draw[gray, thick] (0,0,-0.75)--++(0,0,1.5)--++(0,1,0)--++(0,0,-1.5)--++(0,-1,0);
\draw[gray, thick] (0,0,-0.75)--++(0,0,1.5)--++(0,-1,0)--++(0,0,-1.5)--++(0,1,0);
\draw[gray, thick] (0,0,-0.75)--++(0,0,1.5)--++(1,0,0)--++(0,0,-1.5)--++(-1,0,0)--cycle;
\draw[gray,thick] (0,0,0.75)--++(0,0,0.6);
\draw[fill=red] (0,0,0.75) circle (0.075);
\draw[fill=red]{[canvas is xz plane at y=0] (-0.5,0) circle (0.075)};
\draw[fill=red]{[canvas is xz plane at y=0] (0.5,0) circle (0.075)};
\draw[fill=red]{[canvas is yz plane at x=0] (0.5,0) circle (0.075)};
\draw[fill=red]{[canvas is yz plane at x=0] (-0.5,0) circle (0.075)};
\node[] at (0,0,0) {$X^\alpha_{g}$};
\end{tikzpicture}
, \ 
\begin{tikzpicture}[baseline=0cm]
\draw[densely dotted, red,rounded corners] (-0.5,0,0)--++(0,0.5,0)--++(1,0,0)--++(0,-1,0)--++(-1,0,0)--(-0.5,0,0);
\draw[gray,thick] (0,0,-0.75)--++(0,0,1.5)--++(-1,0,0)--++(0,0,-1.5)--++(+1,0,0);
\draw[gray, thick] (0,0,-0.75)--++(0,0,1.5)--++(0,1,0)--++(0,0,-1.5)--++(0,-1,0);
\draw[gray, thick] (0,0,-0.75)--++(0,0,1.5)--++(0,-1,0)--++(0,0,-1.5)--++(0,1,0);
\draw[gray, thick] (0,0,-0.75)--++(0,0,1.5)--++(1,0,0)--++(0,0,-1.5)--++(-1,0,0)--cycle;
\draw[gray,thick] (0,0,-0.75)--++(0,0,-0.6);
\draw[fill=red] (0,0,-0.75) circle (0.075);
\draw[fill=red]{[canvas is xz plane at y=0] (-0.5,0) circle (0.075)};
\draw[fill=red]{[canvas is xz plane at y=0] (0.5,0) circle (0.075)};
\draw[fill=red]{[canvas is yz plane at x=0] (0.5,0) circle (0.075)};
\draw[fill=red]{[canvas is yz plane at x=0] (-0.5,0) circle (0.075)};
\node[] at (0,0,0) {$X^{\bar{\alpha}}_{g}$};
\end{tikzpicture}
 \ .
$$

Two confined string-like excitations, one constructed with $X^\alpha_{g}$ and the other with $X^{\bar{\alpha}}_{g}$, can be combined to created a deconfined pair of string-like excitations:
$$
\begin{tikzpicture}[baseline=0cm]
\draw[gray, thick] (1,-1,0.75)--++(0,2,0)--++(0,0,-3)--++(0,-2,0)--++(0,0,3);
\draw[gray, thick] (0,-1,0.75)--++(0,2,0)--++(0,0,-3)--++(0,-2,0)--++(0,0,3);
\draw[gray, thick] (-1,0,0.75)--++(3,0,0)--++(0,0,-3)--++(-3,0,0)--++(0,0,3);
\draw[gray, thick] (-1,1,-0.75)--++(3,0,0)--++(0,-2,0)--++(-3,0,0)--++(0,2,0);
\draw[gray, thick] (0,0,0.75) --++(0,0,-3);
\draw[gray, thick] (1,0,0.75) --++(0,0,-3);
\draw[gray, thick] (-1,0,-0.75) --++(3,0,0);
\draw[gray, thick] (0,-1,-0.75) --++(0,2,0);
\draw[gray, thick] (1,-1,-0.75) --++(0,2,0);
\draw[densely dotted, red,rounded corners] (-0.5,0,0)--++(0,0.5,0)--++(2,0,0)--++(0,-1,0)--++(-2,0,0)--(-0.5,0,0);
\draw[fill=red]{[canvas is xz plane at y=0] (-0.5,0) circle (0.075)};
\draw[fill=red]{[canvas is xz plane at y=0] (1.5,0) circle (0.075)};
\draw[fill=red]{[canvas is yz plane at x=0] (0.5,0) circle (0.075)};
\draw[fill=red]{[canvas is yz plane at x=0] (-0.5,0) circle (0.075)};
\draw[fill=red]{[canvas is yz plane at x=1] (0.5,0) circle (0.075)};
\draw[fill=red]{[canvas is yz plane at x=1] (-0.5,0) circle (0.075)};
\draw[fill=red]{[canvas is xz plane at y=0] (-0.5,-1.5) circle (0.075)};
\draw[fill=red]{[canvas is xz plane at y=0] (1.5,-1.5) circle (0.075)};
\draw[fill=red]{[canvas is yz plane at x=0] (0.5,-1.5) circle (0.075)};
\draw[fill=red]{[canvas is yz plane at x=0] (-0.5,-1.5) circle (0.075)};
\draw[fill=red]{[canvas is yz plane at x=1] (0.5,-1.5) circle (0.075)};
\draw[fill=red]{[canvas is yz plane at x=1] (-0.5,-1.5) circle (0.075)};
\draw[densely dotted, red,rounded corners] (-0.5,0,-1.5)--++(0,0.5,0)--++(2,0,0)--++(0,-1,0)--++(-2,0,0)--(-0.5,0,-1.5);
\node[] at (0,0,-1.5) {$X^\alpha_{g}$};
\node[] at (1,0,-1.5) {$X^\alpha_{g}$};
\node[] at (0,0,0) {$X^{\bar{\alpha}}_{g}$};
\node[] at (1,0,0) {$X^{\bar{\alpha}}_{g}$};
\end{tikzpicture}
$$
This combined excitation is a loop of dipoles that can be represented schematically as follows: 
$$
\begin{tikzpicture}
     \draw[thick, red,rounded corners] (0,0.5,-1)--++(0,0.5,0)--++(1,0,0)--++(0,1,0)--++(1,0,0)--++(0,-2,0)--++(-2,0,0)--++(0,0.5,0);
   \draw[thick, red,rounded corners,preaction={draw, line width=2.5pt, white}] (0,0.5,0)--++(0,0.5,0)--++(1,0,0)--++(0,1,0)--++(1,0,0)--++(0,-2,0)--++(-2,0,0)--++(0,0.5,0);
     \draw[fill=cyan, draw=blue, opacity=0.7]{[canvas is zy plane at x=0] (-0.5,0.5) ellipse (0.6 and 0.15)};
     \draw[fill=cyan, draw=blue, opacity=0.7]{[canvas is zy plane at x=0.5] (-0.5,1) ellipse (0.6 and 0.15)};
     \draw[fill=cyan, draw=blue, opacity=0.7]{[canvas is zy plane at x=0.5] (-0.5,0) ellipse (0.6 and 0.15)};
     \draw[fill=cyan, draw=blue, opacity=0.7]{[canvas is zy plane at x=1.5] (-0.5,0) ellipse (0.6 and 0.15)};
     \draw[fill=cyan, draw=blue, opacity=0.7]{[canvas is zy plane at x=2] (-0.5,0.5) ellipse (0.6 and 0.15)};
     \draw[fill=cyan, draw=blue, opacity=0.7]{[canvas is zy plane at x=2] (-0.5,1.5) ellipse (0.6 and 0.15)};
     \draw[fill=cyan, draw=blue, opacity=0.7]{[canvas is zy plane at x=1.5] (-0.5,2) ellipse (0.6 and 0.15)};
          \draw[fill=cyan, draw=blue, opacity=0.7]{[canvas is zy plane at x=1] (-0.5,1.5) ellipse (0.6 and 0.15)};
\end{tikzpicture}
$$

We can also deconfined pairs of loop-like excitations by decorating them with a $Z$-operator:
\begin{equation}\label{memdec}
\begin{tikzpicture}[baseline=0cm]
\draw[densely dotted, red,rounded corners] (-0.5,0,0)--++(0,0.5,0)--++(2,0,0)--++(0,-1,0)--++(-2,0,0)--(-0.5,0,0);
\draw[gray,thick] (0,0,-0.75)--++(0,0,1.5)--++(-1,0,0)--++(0,0,-1.5)--++(+1,0,0);
\draw[gray, thick] (0,0,-0.75)--++(0,0,1.5)--++(0,1,0)--++(0,0,-1.5)--++(0,-1,0);
\draw[gray, thick] (0,0,-0.75)--++(0,0,1.5)--++(0,-1,0)--++(0,0,-1.5)--++(0,1,0);
\draw[gray, thick] (0,0,-0.75)--++(0,0,1.5)--++(1,0,0)--++(0,0,-1.5)--++(-1,0,0)--cycle;
\draw[gray,thick] (0,0,0.75)--++(0,0,0.6);
\draw[gray,thick] (1,0,0.75)--++(0,0,0.6);
\draw[gray, thick] (1,1,-0.75)--++(0,0,1.5)--++(0,-2,0)--++(0,0,-1.5)--++(0,2,0);
\draw[gray, thick] (1,0,-0.75)--++(1,0,0)--++(0,0,1.5)--++(-1,0,0);
\draw[fill=red]{[canvas is xz plane at y=0] (-0.5,0) circle (0.075)};
\draw[fill=red]{[canvas is xz plane at y=0] (1.5,0) circle (0.075)};
\draw[fill=red]{[canvas is yz plane at x=0] (0.5,0) circle (0.075)};
\draw[fill=red]{[canvas is yz plane at x=0] (-0.5,0) circle (0.075)};
\draw[fill=red]{[canvas is yz plane at x=1] (0.5,0) circle (0.075)};
\draw[fill=red]{[canvas is yz plane at x=1] (-0.5,0) circle (0.075)};
\node[] at (0,0,0) {$X^\alpha_{g}$};
\node[] at (0.5,0,0.75) {$Z_{\hat{g}}$};
\node[] at (1,0,0) {$X^\alpha_{g}$};
\end{tikzpicture}  
\end{equation}
This building block of size $1\times 2$, and the analogous $2\times 1$, can be combined in a membrane operator that creates a string-like anyon at its boundary.

As in the regular 3D surface codes, the vertex terms are violated by the $Z_\chi$ operators:
$$
\begin{tikzpicture}[baseline=0cm]
\draw[gray,thick] (0,0,-0.75)--++(0,0,1.5)--++(-1,0,0)--++(0,0,-1.5)--++(+1,0,0);
\draw[gray, thick] (0,0,-0.75)--++(0,0,1.5)--++(0,1,0)--++(0,0,-1.5)--++(0,-1,0);
\draw[gray, thick] (0,0,-0.75)--++(0,0,1.5)--++(0,-1,0)--++(0,0,-1.5)--++(0,1,0);
\draw[gray, thick] (0,0,-0.75)--++(0,0,1.5)--++(1,0,0)--++(0,0,-1.5)--++(-1,0,0)--cycle;
\draw[gray,thick] (0,0,0.75)--++(0,0,0.6);
\draw[gray,thick] (0,0,-0.75)--++(0,0,-0.6);
\draw[fill=red] (0,0,0.75) circle (0.075);
\draw[fill=red] (0,0,-0.75) circle (0.075);
\node[] at (0,0,0) {$Z_{\chi}$};
\end{tikzpicture}
$$
So a string, on the direct lattice, of $Z_\chi$ operators creates two excitations at its ends. These excitations are unconfined and are free to move. They braid non-trivially with membranes created by $X_g$-operators in the $xz$ and $yz$ planes and with the decorated membranes of Eq.~\eqref{memdec} in the $xy$ plane, resulting in a phase factor $\chi(g)$. However, the anyonic $Z_\chi$-string crosses trivially the deconfined doubled membrane of $X^{\bar{\alpha}}_{g}\otimes X^{{\alpha}}_{g}$-operators in the $xy$-plane, that creates the loop of dipoles, since $[Z_\chi\otimes Z_\chi, X^{\bar{\alpha}}_{g}\otimes X^{{\alpha}}_{g}]=0$.

In the regular 3D toric code placed in a 3D torus, there are 3 couples of logical operators associated to uncontractible loops of $Z$-operators and its perpendicular planes formed by $X$ matrices. 

In our twisted model the $xy$-plane operators are constructed by decorated membranes with the building blocks of Eq.~\eqref{memdec}.  Since these have sizes $2\times 1$ and $1\times 2$, the corresponding logical operators cannot be constructed for an odd number of $z$-edges.
Note that the double $xy$-membrane composed of $X^\alpha_g\otimes X^{\bar{\alpha}}_{g}$ is not a logical operator: it is a product of the vertex Hamiltonian terms.

In the same way we could have twisted the edges in the other directions of the vertex Hamiltonian terms, confining the corresponding string-like excitations and decorating the logical operators too.

\subsection{Twisted $\Z_2\times\Z_2 $ X-cube code}

Let us consider the 2-cocycle $\beta$ of the dual of $\Z_2\times\Z_2 $ (the group of its irreps) and we twist the vertical edges of vertex terms ($yz$ and $xy$ planes) of the $\Z_2\times\Z_2 $ X-cube code, resulting in the Hamiltonian:

\begin{align}
H^{\beta}_{\Z_2^2 XC} = & 
-\sum_g 
\begin{tikzpicture}[baseline=0cm]
\draw[ thick, gray] {[canvas is xy plane at z=0] (0,0) rectangle (1,1)};
\draw[ thick, gray,preaction={draw, line width=2.5pt, white}] {[canvas is xy plane at z=1.5] (0,0) rectangle (1,1)};
\draw[ thick, gray] (0,0,0)--++ (0,0,1.5);
\draw[ thick, gray] (1,0,0)--++ (0,0,1.5);
\draw[thick, gray] (1,1,0)--++ (0,0,1.5);
\draw[ thick, gray] (0,1,0)--++ (0,0,1.5);
\node[] at (0,0.5,0) { \footnotesize $X_{g}$};
\node[] at (0.5,1,0) {\footnotesize$X_{g}$};
\node[] at (0.5,0,0) {\footnotesize$X_{g}$};
\node[] at (1,0.5,0) {\footnotesize$X_{g}$};
\node[] at (0,0,0.75) { \footnotesize $X_{g}$};
\node[] at (0,1,0.75) {\footnotesize$X_{g}$};
\node[] at (1,0,0.75) {\footnotesize$X_{g}$};
\node[] at (1,1,0.75) {\footnotesize$X_{g}$};
\node[] at (0,0.5,1.5) {\footnotesize$X_{g}$};
\node[] at (0.5,1,1.5) {\footnotesize$X_{g}$};
\node[] at (0.5,0,1.5) {\footnotesize$X_{g}$};
\node[] at (1,0.5,1.5) {\footnotesize$X_{g}$};
\end{tikzpicture}
\\
& - \sum_\chi
\begin{tikzpicture}[baseline=0cm]
\draw[gray,thick] (0,0,-1.2)--(0,0,1.2);
\draw[gray,thick] (-0.5,0,0)--(0.5,0,0);
\draw[gray,thick] (0,-1,0)--(0,1,0);
\node[] at (0,0,-0.75) {\footnotesize $Z_{\chi}$};
\node[] at (0,0,0.75) {\footnotesize $Z_{\chi}$};
\node[] at (0,0.7,0) {\footnotesize $Z^\beta_{\chi}$};
\node[] at (0,-0.7,0) {\footnotesize $Z^{\bar{\beta}}_{\chi}$};
\end{tikzpicture} 
+
\begin{tikzpicture}[baseline=0cm]
\draw[gray,thick] (0,0,-1.2)--(0,0,1.2);
\draw[gray,thick] (-0.7,0,0)--(0.7,0,0);
\draw[gray,thick] (0,-1,0)--(0,1,0);
\node[] at (0,0.5,0) {\footnotesize$Z^\beta_{\chi}$};
\node[] at (0,-0.5,0) {\footnotesize$Z^{\bar{\beta}}_{\chi}$};
\node[] at (0.35,0,0) {\footnotesize$Z_{\chi}$};
\node[] at (-0.35,0,0) {\footnotesize$Z_{\chi}$};
\end{tikzpicture}
+
\begin{tikzpicture}[baseline=0cm]
\draw[gray,thick] (0,0,-1.5)--(0,0,1.5);
\draw[gray,thick] (-0.7,0,0)--(0.7,0,0);
\draw[gray,thick] (0,-1,0)--(0,1,0);
\node[] at (0,0,-0.85) {\footnotesize$Z_{\chi}$};
\node[] at (0,0,0.85) {\footnotesize$Z_{\chi}$};
\node[] at (0.35,0,0) {\footnotesize$Z_{\chi}$};
\node[] at (-0.35,0,0) {\footnotesize$Z_{\chi}$};
\end{tikzpicture}
\notag
\end{align}
We note that the product of the $yz$-vertex terms in a $yz$-plane results in $X_g$ operators acting on the $y$-edges of that plane (similarly for the $xy$-term).

Let us denote by a red cube the violation of a cube Hamiltonian term (not modified by the twisting). Then a $Z_\chi$ operator on a $\hat{z}$-edge or $\hat{x}$-edge violates its 4 surrounding cube terms. However,  a $Z^{\bar{\beta}}_\chi$ operator in a $\hat{y}$-edge excites its 4 surrounding cube terms plus the 2 vertex terms, corresponding to $yz$ and $xy$ planes, above it. Pictorially:
$$ 
\begin{tikzpicture}[baseline=0cm]
\draw[gray,thick] (0,0,-0.75)--++(0,0,1.5)--++(-1,0,0)--++(0,0,-1.5)--++(+1,0,0);
\draw[gray, thick] (0,0,-0.75)--++(0,0,1.5)--++(0,1,0)--++(0,0,-1.5)--++(0,-1,0);
\draw[gray, thick] (0,0,-0.75)--++(0,0,1.5)--++(0,-1,0)--++(0,0,-1.5)--++(0,1,0);
\draw[gray, thick] (0,0,-0.75)--++(0,0,1.5)--++(1,0,0)--++(0,0,-1.5)--++(-1,0,0)--cycle;
\draw[gray,thick] (0,0,0.75)--++(0,0,0.6);
\draw[gray,thick] (0,0,-0.75)--++(0,0,-0.6);
\pic at (0.5,0.5,0){cube};
\pic at (-0.5,0.5,0){cube};
\pic at (0.5,-0.5,0){cube};
\pic at (-0.5,-0.5,0){cube};
\node[] at (0,0,0) {$Z_{\chi}$};
\end{tikzpicture}
 \ , \ 
\begin{tikzpicture}[baseline=0cm]
\draw[gray,thick] (0,1,0.75)--++(0,0.5,0);
\draw[gray, thick] (0,0,0.75)--++(0,0,-1.5)--++(0,1,0)--++(0,0,1.5)--++(0,-1,0);
\draw[gray, thick] (0,0,0.75)--++(0,0,1.5)--++(0,1,0)--++(0,0,-1.5)--++(0,-1,0);
\pic at (0.5,0.5,0.5){cube};
\pic at (-0.5,0.5,0.5){cube};
\draw[gray, thick] (0,0,0.75)--++(1,0,0)--++(0,1,0)--++(-1,0,0)--++(0,-1,0);
\draw[gray, thick] (0,0,0.75)--++(-1,0,0)--++(0,1,0)--++(1,0,0)--++(0,-1,0);
\draw[fill=red] (0,1,0.75) circle (0.075);
\pic at (0.5,0.5,2){cube};
\pic at (-0.5,0.5,2){cube};
\node[] at (0,0.5,0.75) {$Z^{\bar{\beta}}_{\chi}$};
\end{tikzpicture}
$$
Rectangular combinations of $Z_\chi$-operators on $x$ and $z$ edges creates  four excitations  on the rectangle's edges, which corresponds to the usual fracton with restricted planar mobility of the X-cube code, the so-called planons:

$$
\begin{tikzpicture}
\foreach \y in {0,1,2}{
\foreach \x in {0,1,2}{
\draw[very thick, gray ] {[canvas is xz plane at y=\y] (\x,0) --++(0,1.5)};
\node[] at (\x,\y,0.75) {$Z_{\chi}$};
}}
\pic at (-0.5,2.5,0.75){cube};
\pic at (-0.5,-0.5,0.75){cube};
\pic at (2.5,2.5,0.75){cube};
\pic at (2.5,-0.5,0.75){cube};
\end{tikzpicture}
$$

The extra violations of the vertex terms of the $Z^\beta_{\chi}$ and $Z^{\bar{\beta}}_{\chi}$ operators means that their corresponding generated planons are confined since its energy is proportional to the area of the rectangular operator creating them. However, we can bound both confined planons to construct a planon of dipoles:

$$
\begin{tikzpicture}
\foreach \z in {0,1.5,3}{
\foreach \x in {0,1,2}{
\draw[very thick, gray ] {[canvas is yz plane at x=\x] (-1,\z) --++(2,0)};
\node[] at (\x,0.5,\z) {\footnotesize$Z^\beta_{\chi}$};
\draw[thick,gray ] (\x-0.1,0,\z) --(\x+0.1,0,\z);
\draw[thick,gray ] (\x,0,\z-0.15) --(\x,0,\z+0.15);
\node[] at (\x,-0.5,\z) {\footnotesize$Z^{\bar{\beta}}_{\chi}$};
}}

\draw[fill=cyan, draw=blue, opacity=0.7]{[canvas is xy plane at z=-1](-0.7,0) ellipse (0.15 and 0.4)};
\pic at (-0.7,0.5,-1){cube};
\pic at (-0.7,-0.5,-1){cube};
\draw[fill=cyan, draw=blue, opacity=0.7]{[canvas is xy plane at z=4.25](-0.3,0) ellipse (0.15 and 0.4)};
\pic at (-0.3,0.5,4.25){cube};
\pic at (-0.3,-0.5,4.25){cube};
\draw[fill=cyan, draw=blue, opacity=0.7]{[canvas is xy plane at z=-0.5](2.5,0) ellipse (0.15 and 0.4)};
\pic at (2.5,0.5,-0.5){cube};
\pic at (2.5,-0.5,-0.5){cube};
\draw[fill=cyan, draw=blue, opacity=0.7]{[canvas is xy plane at z=4.25](2.7,0) ellipse (0.15 and 0.4)};
\pic at (2.7,0.5,4.25){cube};
\pic at (2.7,-0.5,4.25){cube};
\end{tikzpicture}
$$
This type of excitation has the same restricted mobility of the planon. 
The other mechanism to unconfine is the decoration with $X$-operators in a $xz$-plaquette as follows:

$$
\begin{tikzpicture}
\foreach \z in {0,3}{
\foreach \x in {0,2}{
\draw[very thick, gray ] (\x,0,-0.4) --++(0,0,3.8);
\draw[very thick, gray ] (\x,0.3,\z) --++(0,-1,0);
\node[] at (\x,-0.4,\z) {$Z^{\bar{\beta}}_{\hat{g}}$};
\node[] at (\x,0,1.5) {$X_g$};
}
\draw[very thick, gray ] (-0.3,0,\z) --++(2.6,0,0);
\node[] at (1,0,\z) {$X_g$};}

\pic at (-1.5,-0.75,-1.75){cube};
\pic at (2,-0.75,-1.75){cube};
\pic at (-1.2,-0.75,2.25){cube};
\pic at (2.25,-0.75,2.25){cube};
\end{tikzpicture}
$$
The regular $X_g$-strings creating point-like excitations at its ends, which are also deconfined on this model, braid trivially with the planon-like dipoles but non-trivially with the decorated planons.

\section{Conclusion and Outlook}

In this work, we introduced a family of Hamiltonian models that realize novel patterns of anyon and fracton confinement, resulting in a system-size-dependent ground state degeneracy. We identified two mechanisms for partial deconfinement: the formation of dipolar bound states from confined excitations, and the decoration of these excitations with oppositely charged anyons to construct dyon-like, deconfined excitations.

A distinctive feature of our construction is the explicit and directional nature of the confinement, which, unlike traditional mechanisms in topological phases, does not rely on anyon condensation. Instead, confinement is engineered directly through Hamiltonian modifications that break spatial symmetries, giving rise to anisotropic mobility of excitations. This raises compelling questions about the possibility of interpolating between the standard quantum double phase and the confined phase described here, potentially via local perturbations—such as those breaking translation invariance.

Notably, partial confinement has also been explored as a mechanism for fracton emergence, as discussed in Ref.~\cite{Ma_2018}. Given that one of our 2D models exhibits fractal-like excitations, it is natural to ask whether a similar confinement-based modification of the 3D surface code could yield fracton phases in a more direct and constructive manner.

Our results also underscore the sensitivity of the confinement mechanism to the geometry of the lattice and the specific Hamiltonian terms involved. In particular, only interaction terms with certain overlapping structures can be consistently twisted. This indicates that the resulting order is not purely topological, but rather depends intricately on microscopic and symmetry-breaking details.

From an experimental or diagnostic standpoint, the directional confinement introduced in these models could be probed using direction-dependent Fredenhagen-Marcu order parameters \cite{Gregor_2011}. Such observables would serve as clear signatures of anisotropic confinement and could help distinguish these models from conventional topological phases.

Finally, the emergence of unpaired logical operators in certain regimes suggests the intriguing possibility of storing classical information with topological protection. Exploring whether this can be developed into a robust hybrid classical-quantum code remains an interesting direction for future work.

\section*{Acknowledgments}
I thank J.~Molina-Vilaplana for useful discussions on a related project, and Anasuya Lyons for helpful comments.
The author is funded by the FWF Erwin Schrödinger Program (Grant DOI 10.55776/J4796).

\bibliography{bibliography}

\begin{thebibliography}{21}%
\makeatletter
\providecommand \@ifxundefined [1]{%
 \@ifx{#1\undefined}
}%
\providecommand \@ifnum [1]{%
 \ifnum #1\expandafter \@firstoftwo
 \else \expandafter \@secondoftwo
 \fi
}%
\providecommand \@ifx [1]{%
 \ifx #1\expandafter \@firstoftwo
 \else \expandafter \@secondoftwo
 \fi
}%
\providecommand \natexlab [1]{#1}%
\providecommand \enquote  [1]{``#1''}%
\providecommand \bibnamefont  [1]{#1}%
\providecommand \bibfnamefont [1]{#1}%
\providecommand \citenamefont [1]{#1}%
\providecommand \href@noop [0]{\@secondoftwo}%
\providecommand \href [0]{\begingroup \@sanitize@url \@href}%
\providecommand \@href[1]{\@@startlink{#1}\@@href}%
\providecommand \@@href[1]{\endgroup#1\@@endlink}%
\providecommand \@sanitize@url [0]{\catcode `\\12\catcode `\$12\catcode `\&12\catcode `\#12\catcode `\^12\catcode `\_12\catcode `\%12\relax}%
\providecommand \@@startlink[1]{}%
\providecommand \@@endlink[0]{}%
\providecommand \url  [0]{\begingroup\@sanitize@url \@url }%
\providecommand \@url [1]{\endgroup\@href {#1}{\urlprefix }}%
\providecommand \urlprefix  [0]{URL }%
\providecommand \Eprint [0]{\href }%
\providecommand \doibase [0]{https://doi.org/}%
\providecommand \selectlanguage [0]{\@gobble}%
\providecommand \bibinfo  [0]{\@secondoftwo}%
\providecommand \bibfield  [0]{\@secondoftwo}%
\providecommand \translation [1]{[#1]}%
\providecommand \BibitemOpen [0]{}%
\providecommand \bibitemStop [0]{}%
\providecommand \bibitemNoStop [0]{.\EOS\space}%
\providecommand \EOS [0]{\spacefactor3000\relax}%
\providecommand \BibitemShut  [1]{\csname bibitem#1\endcsname}%
\let\auto@bib@innerbib\@empty
\bibitem [{\citenamefont {Kitaev}(2003)}]{Kitaev03}%
  \BibitemOpen
  \bibfield  {author} {\bibinfo {author} {\bibfnamefont {A.}~\bibnamefont {Kitaev}},\ }\href {https://doi.org/https://doi.org/10.1016/S0003-4916(02)00018-0} {\bibfield  {journal} {\bibinfo  {journal} {Annals of Physics}\ }\textbf {\bibinfo {volume} {303}},\ \bibinfo {pages} {2 } (\bibinfo {year} {2003})}\BibitemShut {NoStop}%
\bibitem [{\citenamefont {Preskill}(2004)}]{Preskill04}%
  \BibitemOpen
  \bibfield  {author} {\bibinfo {author} {\bibfnamefont {J.}~\bibnamefont {Preskill}},\ }\href@noop {} {\emph {\bibinfo {title} {Chapter 9. Topological Quantum Computation}}}\ (\bibinfo  {publisher} {California Institute of Technology},\ \bibinfo {year} {2004})\BibitemShut {NoStop}%
\bibitem [{\citenamefont {Bais}\ \emph {et~al.}(2002)\citenamefont {Bais}, \citenamefont {Schroers},\ and\ \citenamefont {Slingerland}}]{Bais02}%
  \BibitemOpen
  \bibfield  {author} {\bibinfo {author} {\bibfnamefont {F.~A.}\ \bibnamefont {Bais}}, \bibinfo {author} {\bibfnamefont {B.~J.}\ \bibnamefont {Schroers}},\ and\ \bibinfo {author} {\bibfnamefont {J.~K.}\ \bibnamefont {Slingerland}},\ }\href {https://doi.org/10.1103/PhysRevLett.89.181601} {\bibfield  {journal} {\bibinfo  {journal} {Phys. Rev. Lett.}\ }\textbf {\bibinfo {volume} {89}},\ \bibinfo {pages} {181601} (\bibinfo {year} {2002})}\BibitemShut {NoStop}%
\bibitem [{\citenamefont {Vidal}\ \emph {et~al.}(2009)\citenamefont {Vidal}, \citenamefont {Dusuel},\ and\ \citenamefont {Schmidt}}]{Vidal_2009}%
  \BibitemOpen
  \bibfield  {author} {\bibinfo {author} {\bibfnamefont {J.}~\bibnamefont {Vidal}}, \bibinfo {author} {\bibfnamefont {S.}~\bibnamefont {Dusuel}},\ and\ \bibinfo {author} {\bibfnamefont {K.~P.}\ \bibnamefont {Schmidt}},\ }\bibfield  {journal} {\bibinfo  {journal} {Physical Review B}\ }\textbf {\bibinfo {volume} {79}},\ \href {https://doi.org/10.1103/physrevb.79.033109} {10.1103/physrevb.79.033109} (\bibinfo {year} {2009})\BibitemShut {NoStop}%
\bibitem [{\citenamefont {Iqbal}\ and\ \citenamefont {Schuch}(2021)}]{IqbalSuch21}%
  \BibitemOpen
  \bibfield  {author} {\bibinfo {author} {\bibfnamefont {M.}~\bibnamefont {Iqbal}}\ and\ \bibinfo {author} {\bibfnamefont {N.}~\bibnamefont {Schuch}},\ }\href {https://doi.org/10.1103/PhysRevX.11.041014} {\bibfield  {journal} {\bibinfo  {journal} {Phys. Rev. X}\ }\textbf {\bibinfo {volume} {11}},\ \bibinfo {pages} {041014} (\bibinfo {year} {2021})}\BibitemShut {NoStop}%
\bibitem [{\citenamefont {Vijay}\ \emph {et~al.}(2016)\citenamefont {Vijay}, \citenamefont {Haah},\ and\ \citenamefont {Fu}}]{Vijay16}%
  \BibitemOpen
  \bibfield  {author} {\bibinfo {author} {\bibfnamefont {S.}~\bibnamefont {Vijay}}, \bibinfo {author} {\bibfnamefont {J.}~\bibnamefont {Haah}},\ and\ \bibinfo {author} {\bibfnamefont {L.}~\bibnamefont {Fu}},\ }\bibfield  {journal} {\bibinfo  {journal} {Physical Review B}\ }\textbf {\bibinfo {volume} {94}},\ \href {https://doi.org/10.1103/physrevb.94.235157} {10.1103/physrevb.94.235157} (\bibinfo {year} {2016})\BibitemShut {NoStop}%
\bibitem [{\citenamefont {Brown}\ and\ \citenamefont {Williamson}(2020)}]{Brown20}%
  \BibitemOpen
  \bibfield  {author} {\bibinfo {author} {\bibfnamefont {B.~J.}\ \bibnamefont {Brown}}\ and\ \bibinfo {author} {\bibfnamefont {D.~J.}\ \bibnamefont {Williamson}},\ }\bibfield  {journal} {\bibinfo  {journal} {Physical Review Research}\ }\textbf {\bibinfo {volume} {2}},\ \href {https://doi.org/10.1103/physrevresearch.2.013303} {10.1103/physrevresearch.2.013303} (\bibinfo {year} {2020})\BibitemShut {NoStop}%
\bibitem [{\citenamefont {Song}\ \emph {et~al.}(2022)\citenamefont {Song}, \citenamefont {Sch\"onmeier-Kromer}, \citenamefont {Liu}, \citenamefont {Viyuela}, \citenamefont {Pollet},\ and\ \citenamefont {Martin-Delgado}}]{Song22}%
  \BibitemOpen
  \bibfield  {author} {\bibinfo {author} {\bibfnamefont {H.}~\bibnamefont {Song}}, \bibinfo {author} {\bibfnamefont {J.}~\bibnamefont {Sch\"onmeier-Kromer}}, \bibinfo {author} {\bibfnamefont {K.}~\bibnamefont {Liu}}, \bibinfo {author} {\bibfnamefont {O.}~\bibnamefont {Viyuela}}, \bibinfo {author} {\bibfnamefont {L.}~\bibnamefont {Pollet}},\ and\ \bibinfo {author} {\bibfnamefont {M.~A.}\ \bibnamefont {Martin-Delgado}},\ }\href {https://doi.org/10.1103/PhysRevLett.129.230502} {\bibfield  {journal} {\bibinfo  {journal} {Phys. Rev. Lett.}\ }\textbf {\bibinfo {volume} {129}},\ \bibinfo {pages} {230502} (\bibinfo {year} {2022})}\BibitemShut {NoStop}%
\bibitem [{\citenamefont {Kitaev}\ and\ \citenamefont {Kong}(2012)}]{Kitaev12}%
  \BibitemOpen
  \bibfield  {author} {\bibinfo {author} {\bibfnamefont {A.}~\bibnamefont {Kitaev}}\ and\ \bibinfo {author} {\bibfnamefont {L.}~\bibnamefont {Kong}},\ }\bibfield  {journal} {\bibinfo  {journal} {Commun. Math. Phys}\ }\href {https://doi.org/doi.org/10.1007/s00220-012-1500-5} {doi.org/10.1007/s00220-012-1500-5} (\bibinfo {year} {2012})\BibitemShut {NoStop}%
\bibitem [{\citenamefont {Chen}\ \emph {et~al.}(2025)\citenamefont {Chen}, \citenamefont {Liu}, \citenamefont {Zhang}, \citenamefont {Liang}, \citenamefont {Chen}, \citenamefont {Liu},\ and\ \citenamefont {Song}}]{chen2025}%
  \BibitemOpen
  \bibfield  {author} {\bibinfo {author} {\bibfnamefont {K.}~\bibnamefont {Chen}}, \bibinfo {author} {\bibfnamefont {Y.}~\bibnamefont {Liu}}, \bibinfo {author} {\bibfnamefont {Y.}~\bibnamefont {Zhang}}, \bibinfo {author} {\bibfnamefont {Z.}~\bibnamefont {Liang}}, \bibinfo {author} {\bibfnamefont {Y.-A.}\ \bibnamefont {Chen}}, \bibinfo {author} {\bibfnamefont {K.}~\bibnamefont {Liu}},\ and\ \bibinfo {author} {\bibfnamefont {H.}~\bibnamefont {Song}},\ }\href {https://arxiv.org/abs/2503.04699} {\bibinfo {title} {Anyon theory and topological frustration of high-efficiency quantum ldpc codes}} (\bibinfo {year} {2025}),\ \Eprint {https://arxiv.org/abs/2503.04699} {arXiv:2503.04699 [quant-ph]} \BibitemShut {NoStop}%
\bibitem [{Note1()}]{Note1}%
  \BibitemOpen
  \bibinfo {note} {The set of 2-cocycles are classified by the second cohomology group, $\protect \mathcal {H}^2[G,U(1)]$, which is abelian.}\BibitemShut {Stop}%
\bibitem [{\citenamefont {Garre-Rubio}(2024)}]{Garre24Emergent}%
  \BibitemOpen
  \bibfield  {author} {\bibinfo {author} {\bibfnamefont {J.}~\bibnamefont {Garre-Rubio}},\ }\href {https://doi.org/10.1038/s41467-024-52320-7} {\bibfield  {journal} {\bibinfo  {journal} {Nat. Commun.}\ }\textbf {\bibinfo {volume} {15}} (\bibinfo {year} {2024})}\BibitemShut {NoStop}%
\bibitem [{\citenamefont {Bombin}(2010)}]{Bombin_2010}%
  \BibitemOpen
  \bibfield  {author} {\bibinfo {author} {\bibfnamefont {H.}~\bibnamefont {Bombin}},\ }\bibfield  {journal} {\bibinfo  {journal} {Physical Review Letters}\ }\textbf {\bibinfo {volume} {105}},\ \href {https://doi.org/10.1103/physrevlett.105.030403} {10.1103/physrevlett.105.030403} (\bibinfo {year} {2010})\BibitemShut {NoStop}%
\bibitem [{\citenamefont {Verstraete}\ and\ \citenamefont {Cirac}()}]{Verstraete04}%
  \BibitemOpen
  \bibfield  {author} {\bibinfo {author} {\bibfnamefont {F.}~\bibnamefont {Verstraete}}\ and\ \bibinfo {author} {\bibfnamefont {J.~I.}\ \bibnamefont {Cirac}},\ }\href@noop {} {\bibinfo  {journal} {ArXiv: cond-mat/0407066}\ }\BibitemShut {NoStop}%
\bibitem [{\citenamefont {{\c{S}}ahino{\u{g}}lu}\ \emph {et~al.}(2021)\citenamefont {{\c{S}}ahino{\u{g}}lu}, \citenamefont {Williamson}, \citenamefont {Bultinck}, \citenamefont {Mariën}, \citenamefont {Haegeman}, \citenamefont {Schuch},\ and\ \citenamefont {Verstraete}}]{Sahinoglu14}%
  \BibitemOpen
\bibfield  {journal} {  }\bibfield  {author} {\bibinfo {author} {\bibfnamefont {M.~B.}\ \bibnamefont {{\c{S}}ahino{\u{g}}lu}}, \bibinfo {author} {\bibfnamefont {D.}~\bibnamefont {Williamson}}, \bibinfo {author} {\bibfnamefont {N.}~\bibnamefont {Bultinck}}, \bibinfo {author} {\bibfnamefont {M.}~\bibnamefont {Mariën}}, \bibinfo {author} {\bibfnamefont {J.}~\bibnamefont {Haegeman}}, \bibinfo {author} {\bibfnamefont {N.}~\bibnamefont {Schuch}},\ and\ \bibinfo {author} {\bibfnamefont {F.}~\bibnamefont {Verstraete}},\ }\href@noop {} {\bibfield  {journal} {\bibinfo  {journal} {Annales Henri Poincar{\'{e}}}\ }\textbf {\bibinfo {volume} {22}},\ \bibinfo {pages} {563} (\bibinfo {year} {2021})}\BibitemShut {NoStop}%
\bibitem [{\citenamefont {Schuch}\ \emph {et~al.}(2010)\citenamefont {Schuch}, \citenamefont {Cirac},\ and\ \citenamefont {Perez-Garcia}}]{Schuch10}%
  \BibitemOpen
  \bibfield  {author} {\bibinfo {author} {\bibfnamefont {N.}~\bibnamefont {Schuch}}, \bibinfo {author} {\bibfnamefont {I.}~\bibnamefont {Cirac}},\ and\ \bibinfo {author} {\bibfnamefont {D.}~\bibnamefont {Perez-Garcia}},\ }\href {https://doi.org/https://doi.org/10.1016/j.aop.2010.05.008} {\bibfield  {journal} {\bibinfo  {journal} {Annals of Physics}\ }\textbf {\bibinfo {volume} {325}},\ \bibinfo {pages} {2153 } (\bibinfo {year} {2010})}\BibitemShut {NoStop}%
\bibitem [{\citenamefont {Newman}\ and\ \citenamefont {Moore}(1999)}]{Newman_1999}%
  \BibitemOpen
  \bibfield  {author} {\bibinfo {author} {\bibfnamefont {M.~E.~J.}\ \bibnamefont {Newman}}\ and\ \bibinfo {author} {\bibfnamefont {C.}~\bibnamefont {Moore}},\ }\href {https://doi.org/10.1103/physreve.60.5068} {\bibfield  {journal} {\bibinfo  {journal} {Physical Review E}\ }\textbf {\bibinfo {volume} {60}},\ \bibinfo {pages} {5068–5072} (\bibinfo {year} {1999})}\BibitemShut {NoStop}%
\bibitem [{\citenamefont {Zhu}\ \emph {et~al.}(2022)\citenamefont {Zhu}, \citenamefont {Jochym-O'Connor},\ and\ \citenamefont {Dua}}]{Zhu-nofraclog22}%
  \BibitemOpen
  \bibfield  {author} {\bibinfo {author} {\bibfnamefont {G.}~\bibnamefont {Zhu}}, \bibinfo {author} {\bibfnamefont {T.}~\bibnamefont {Jochym-O'Connor}},\ and\ \bibinfo {author} {\bibfnamefont {A.}~\bibnamefont {Dua}},\ }\href {https://doi.org/10.1103/PRXQuantum.3.030338} {\bibfield  {journal} {\bibinfo  {journal} {PRX Quantum}\ }\textbf {\bibinfo {volume} {3}},\ \bibinfo {pages} {030338} (\bibinfo {year} {2022})}\BibitemShut {NoStop}%
\bibitem [{\citenamefont {Blanik}\ \emph {et~al.}(2025)\citenamefont {Blanik}, \citenamefont {Garre-Rubio},\ and\ \citenamefont {Schuch}}]{blanik25}%
  \BibitemOpen
  \bibfield  {author} {\bibinfo {author} {\bibfnamefont {D.}~\bibnamefont {Blanik}}, \bibinfo {author} {\bibfnamefont {J.}~\bibnamefont {Garre-Rubio}},\ and\ \bibinfo {author} {\bibfnamefont {N.}~\bibnamefont {Schuch}},\ }\href {https://arxiv.org/abs/2504.14380} {\bibinfo {title} {Gauging quantum phases: A matrix product state approach}} (\bibinfo {year} {2025}),\ \Eprint {https://arxiv.org/abs/2504.14380} {arXiv:2504.14380 [quant-ph]} \BibitemShut {NoStop}%
\bibitem [{\citenamefont {Ma}\ \emph {et~al.}(2018)\citenamefont {Ma}, \citenamefont {Hermele},\ and\ \citenamefont {Chen}}]{Ma_2018}%
  \BibitemOpen
  \bibfield  {author} {\bibinfo {author} {\bibfnamefont {H.}~\bibnamefont {Ma}}, \bibinfo {author} {\bibfnamefont {M.}~\bibnamefont {Hermele}},\ and\ \bibinfo {author} {\bibfnamefont {X.}~\bibnamefont {Chen}},\ }\bibfield  {journal} {\bibinfo  {journal} {Physical Review B}\ }\textbf {\bibinfo {volume} {98}},\ \href {https://doi.org/10.1103/physrevb.98.035111} {10.1103/physrevb.98.035111} (\bibinfo {year} {2018})\BibitemShut {NoStop}%
\bibitem [{\citenamefont {Gregor}\ \emph {et~al.}(2011)\citenamefont {Gregor}, \citenamefont {Huse}, \citenamefont {Moessner},\ and\ \citenamefont {Sondhi}}]{Gregor_2011}%
  \BibitemOpen
  \bibfield  {author} {\bibinfo {author} {\bibfnamefont {K.}~\bibnamefont {Gregor}}, \bibinfo {author} {\bibfnamefont {D.~A.}\ \bibnamefont {Huse}}, \bibinfo {author} {\bibfnamefont {R.}~\bibnamefont {Moessner}},\ and\ \bibinfo {author} {\bibfnamefont {S.~L.}\ \bibnamefont {Sondhi}},\ }\href {https://doi.org/10.1088/1367-2630/13/2/025009} {\bibfield  {journal} {\bibinfo  {journal} {New Journal of Physics}\ }\textbf {\bibinfo {volume} {13}},\ \bibinfo {pages} {025009} (\bibinfo {year} {2011})}\BibitemShut {NoStop}%
\end{thebibliography}%
\end{document}